\documentclass[bibyear]{aa}  
\usepackage{cuted, ragged2e}
\usepackage{natbib}
\bibpunct{(}{)}{;}{a}{}{,} 
\usepackage{graphicx}
\graphicspath{ {./images/} }

\usepackage{txfonts}
\usepackage{amsmath}
\usepackage{mathtools}
\usepackage{afterpage}

\usepackage[normalem]{ulem}
\usepackage{color}

\begin{document}

   \title{[CII] line intensity mapping the epoch of reionization with the 
   Prime-Cam on FYST}

   \subtitle{I. Line intensity mapping predictions using the Illustris TNG hydrodynamical simulation}

   \author{C. Karoumpis
          \inst{1}
          \and
          B. Magnelli\inst{1}
          \and
          E.Romano-D{\'\i}az \inst{1}
          \and
          M. Haslbauer \inst{2,3}
          \and
          F. Bertoldi \inst{1}}
          
   \institute{Argelander Institut für Astronomie, Universität Bonn,
              Auf dem Hügel 71, D-53121 Bonn, Germany\\
             \email{karoumpis@astro.uni-bonn.de}
             \and
             Helmholtz-Institut f\"ur Strahlen- und Kernphysik, University of Bonn, Nussallee 14-16, D-53115 Bonn, Germany 
              \and
             Max-Planck-Institut f\"ur Radioastronomie, Auf dem H\"ugel 69, D-53121 Bonn, Germany
             }

  \abstract
{} 
{We predict the three-dimensional intensity power spectrum (PS) of the [CII] 158$\,\mu$m line throughout the epoch of (and post) reionization at redshifts from $\approx$ 3.5 to 8. We study the detectability of the PS in a line intensity mapping (LIM) survey with the Prime-Cam spectral-imager on the Fred Young Submillimeter Telescope (FYST).}
{We created mock [CII] tomographic scans in redshift bins at $z\approx 3.7$, 4.3, 5.8, and 7.4 using the Illustris TNG300-1 $\Lambda$CDM simulation and adopting a relation between the star formation activity and the [CII] luminosity ($L_{\rm [CII]}$) of galaxies. A star formation rate (SFR) was assigned to a dark matter halo in the Illustris simulation in two ways: (i) we adopted the SFR computed in the Illustris simulation and, (ii) we matched the abundance of the halos with the SFR traced by the observed dust-corrected ultraviolet luminosity function of high-redshift galaxies. The $L_{\rm [CII]}$ is related to the SFR from a semi-analytic model of galaxy formation, from a hydrodynamical simulation of a high-redshift galaxy, or from a high-redshift [CII] galaxy survey. The [CII] intensity PS was computed from mock tomographic scans to assess its detectability with the anticipated observational capability of the FYST.}
{The amplitude of the predicted [CII] intensity power spectrum varies by more than a factor of 10, depending on the choice of the halo-to-galaxy SFR and the SFR-to-$L_{\rm [CII]}$ relations. 
In the planned $4^{\circ} \times 4^{\circ}$ FYST LIM survey, we expect a detection of the [CII] PS up to $z \approx 5.8$, and potentially even up to $z \approx 7.4$. The design of the envisioned FYST LIM survey enables a PS measurement not only in small (<10 Mpc) shot noise-dominated scales, but also in large (> 50 Mpc) clustering-dominated scales making it the first LIM experiment that will place constraints on the SFR-to-$L_{\rm [CII]}$ and the halo-to-galaxy SFR relations simultaneously.}
{}


   \maketitle
%

\section{Introduction} \label{Introduction}

The epoch of reionization (EoR), which took place 300-1000 million years after the Big Bang, was a period of radical changes in the Universe. The appearance of the first proto-galaxies was followed by the second known hydrogen phase transition--after the one that occurred during recombination--this time from neutral to fully ionized. What triggered this transition and its relation to the appearance of the first light sources is still under investigation. The star formation in galaxies, especially those with low stellar mass, is assumed to be the primary energy source behind the reionization \citep[e.g.,][]{Zaroubi2013,McQuinn2016}, even though there have been many debates on the role of the active galactic nuclei (AGN) as alternative energy sources \cite[e.g.,][]{Haardt2015, Kulkarni2017, Chardin2017, Hassan2018, Mitra2018, Garaldi2019}. The EoR was followed by a period of rapid galaxy maturation (1-1.5 million years after the Big Bang, from now on post-EoR) when galaxies build up their stellar mass and the rest of the characteristic properties (e.g., gas and dust mass, star formation rate), approaching those observed in local galaxies. Although there is a consensus picture developing on the observational characteristics (e.g., rest-frame UV, IR, submillimeter, and radio luminosity) of post-EoR galaxies \citep[e.g.,][]{Madau:Dickinson2014}, most of those are still biased toward the brightest objects, with the few exceptions that already seem to report tensions with the current galaxy evolution models \cite[e.g.,][]{Yang2019,Breysse2021}.

Detecting the faintest EoR galaxies that, according to theoretical models, \citep[e.g.,][]{Kuhlen} powered the reionization of the Universe, as well as their immediate descendants, is one of the next challenges in observational astronomy. Although they are large in numbers, these faint sources might not be easily detectable by traditional galaxy surveys, even in those performed by the most powerful telescopes to date or the coming {\em James Webb Space Telescope} \citep[\textit{JWST};][]{Robertson2013,JWST2019}. An answer to this observational challenge might instead come from the line intensity mapping (LIM) technique, a method extensively used in 21-cm fluctuation research and currently utilized by several new projects focusing on different atomic and molecular lines \citep{Kovetz}. LIM measures the integrated emission of spectral lines from galaxies, with its data products being three-dimensional tomographic scans for which line-of-sight distance information is directly acquired through its frequency dependence. As all photons coming from the targeted angle and line frequency are measured, that is, including those from galaxies too faint to be individually detected by any traditional galaxy survey, LIM of infrared lines can trace the total cosmic star formation rate density (SFRD) during the (post-)EoR, allowing thereby to estimate the total contribution of star-forming (SF) galaxies to the energy budget of the reionization. \cite[e.g.,][]{Basu2004,Righi2008,Lidz:Furlanetto2011,Gong:Cooray:Silva2011,Gong:Cooray:2012,Gong:Cooray:Silva2017,Breysse14,Mashian2015,Yue:Ferrara2015,Yue:Ferrara2019,Lidz:Taylor2016,Comaschi2016,Kovetz,Dumitru2019,Padmanabhan2019,Chung:Viero2018}. Additionally, due to its low angular and spectral resolution requirements, it can be performed over a large portion of the sky by dedicated, small-aperture, wide field of view instruments. This way, LIM will also trace the incipient large-scale structure of the matter distribution in the early Universe, which otherwise would be inaccessible to deep pencil-beam-like observations from the {\em JWST}. Finally, by cross-correlating maps of lines that trace different phases of the interstellar and intergalactic medium, LIM can provide a thorough picture of the phase transition of the intergalactic medium during the \ \ (post-)EoR \citep[e.g.,][]{Gong:Cooray:2012,Dumitru2019}.

The astrophysical information will be extracted from the intensity maps using a two-point statistics, more in concrete the power spectrum. The comoving three-dimensional spherically averaged power spectrum (PS) of the tomographic scans will be the primary measurement of the upcoming first generation of LIM experiments. Not only is PS a promising tool for constraining astrophysical parameters, but it also is the most feasible measurement as well: the mean intensity measurement will be challenged by the smoothness of the FIR continuum foreground emission, while the accuracy of higher-order statistics measurements will be seriously limited by their sample variance.

Among all the possible lines emitted by high-redshift galaxies, the fine structure line of single ionized carbon, [CII], at rest-frame 157.7$\,\mu$m is of particular interest. Indeed, because the [CII] line is a dominant coolant in different phases of the interstellar medium \citep{Carilli:Walter2013}, it is the brightest line in typical SF galaxies and corresponds to about 0.1--1\% of their total bolometric luminosity \citep{Crawford1985,Stacey1991,Wright1991,Lord1996}. It has excellent potential as a dust-unbiased probe of the interstellar gas and the star formation activity of galaxies \citep[e.g.,][]{DeLooze}. Finally, it is an ideal target for ground-based telescopes because its observed frequencies for galaxies at $z\approx4-10$ lie in the submillimeter (submm) atmospheric windows.

In the coming decade, several new experiments will try to trace the evolution of the [CII] PS during the (post-)EoR. Examples include the Tomographic Intensity Mapping Experiment \citep[TIME;][]{TIME} mounted on the Arizona Radio Observatory (ARO) 12 meters radio telescope and the CarbON CII line in post-rEionisation and ReionisaTiOn epoch \citep[CONCERTO;][]{Serra2017,Lagache2018} mounted on the Atacama Pathfinder EXperiment (APEX) 12 meters radio telescope. In this paper, we focus our analysis on the capabilities of the Prime-Cam spectro-imager that will be mounted on the Fred Young Submillimeter Telescope \citep[FYST\footnote{http://www.ccatobservatory.org};][]{Stacey,vavagiakis/etal:2018,choi/etal:2019}. FYST is a 6-meters diameter telescope with a large field of view, while the Prime-Cam spectro-imager will include a state-of-the-art Fabry-Pero (sub)millimetric spectrometer. FYST will be located at a high altitude (5600m) on Cerro Chajnantor, 600 meters above the ALMA plateau. With its large field of view and exquisite location, FYST will efficiently perform large (sub)mm surveys focusing primarily on intensity mapping and cosmic microwave background observations.

To test the feasibility of the [CII] LIM experiment with FYST as well as to optimize its survey strategy, accurate predictions of the [CII] PS at the (post-)EoR are now urgently needed. To make such predictions in the $\Lambda$CDM framework, one usually starts from a mock dark matter (DM) halo lightcone catalog. Then, based on the DM halo properties, one predicts the star formation rate (SFR) of galaxies occupying them, either by using physical or empirical models. The former couple semi-analytic models of galaxy formation to DM-only simulations \citep{Gong:Cooray:2012,Dumitru2019}. The later match observations of UV galaxy surveys with analytic predictions of the DM halo mass function or halo catalogs coming from DM-only simulations, for example, by using the technique called abundance matching \citep[e.g.,][]{Yue:Ferrara2015,Chung:Viero2018}. Alternatively, there are also empirical models which, instead of using high-redshift observations, are calibrated using the cosmic infrared background \citep{Serra2017} or [CII] surveys of local galaxies \citep{Padmanabhan2019}. The different methods of assigning SFRs to galaxies in those DM halos can lead to a factor of four differences in [CII] PS forecasts \citep{Chung:Viero2018}.

The next step is to translate the SFR of each galaxy into a [CII] line luminosity ($L_{\rm [CII]}$) using a SFR-to-$L_{\rm [CII]}$ relation. Despite the complex emission mechanisms, the [CII] emission strongly correlates with the SFR of galaxies in the local Universe. However, for galaxies at higher redshifts, we have to rely on much sparser empirical constraints. Thus, previous works on high-redshift [CII] PS have either used scaling relations based on local Universe observations \citep[e.g.,][]{Spinoglio,DeLooze,Herrera} or simulations of high-redshift galaxies \citep[e.g.,][]{Vallini2015}. The different SFR-to-$L_{[\rm CII]}$ scaling relations can lead to up to two orders of magnitude differences in the predicted [CII] PS \citep{Yue:Ferrara2019}. 

The combination of these two very uncertain steps seems to result in more than two orders of magnitude inconsistencies in the [CII] PS predictions found in the literature \citep{Gong:Cooray:2012,Silva2015,Serra2017,Chung:Viero2018,Dumitru2019,Padmanabhan2019}. Still, a consistent comparison between these forecasts is challenging because they are based on different assumptions and cosmological models (e.g., different spatial and mass resolution, ways of generating the DM-halo catalog). These discrepancies hamper the design of LIM surveys and prevent us from assessing realistically the constraints one will obtain from LIM experiments. In this paper, we produce a set of alternative [CII] PS predictions to study their differences coherently, allowing the optimization of the FYST LIM survey strategy. To do so, we start from a common dark matter cone built using the DM halo catalog of the Illustris TNG300-1 (from now on TNG300-1) hydrodynamical simulation. Subsequently, we apply different models of occupying the DM halos with mock galaxies as well as various SFR-to-$L_{\rm [CII]}$ coupling relations. Finally, from all these versions of mock [CII] tomographic scans, we predict the range of possible values probed by the [CII] PS during the (post-)EoR and study the detectability of this signal by the Prime-cam spectro-imager mounted on the FYST.

Our paper is organized as follows. In Section \ref{Methods}, we present our DM halo lightcone catalog, the two different methods used to populate its DM halos with galaxies (hydrodynamical simulation and abundance matching), and the three different SFR-$L_{\rm{[CII]}}$ scaling relations used in this paper. In Section \ref{Results}, we present our predictions of the [CII] mean intensity and [CII] PS at various redshifts and study the detectability of the PS with the FYST LIM survey. Finally, in Section \ref{discussion} we discuss our results in the context of other state-of-the-art observatories targeting \\ (post-)EoR galaxies. 

This paper is part of a two-paper series. In the second paper, we will investigate ways of removing foreground contamination of astrophysical origin like the cosmic infrared background and carbon monoxide lines coming from lower redshift galaxies \citep{Cheng:Chang:Bock:2016,Sun2018}.

We assume the same $\Lambda$ cold dark matter cosmology as in Illustris TNG: $\Omega _{\rm{m}}= \Omega _{\rm{dm}}+ \Omega _{\rm{b}}=0.3089$, $\Omega _{\rm{b}} =0.0486$, $\Omega_{\rm{\Lambda}}=0.6911$, $\sigma_{8}=0.8159$, $n_{\rm s}=0.9667$, and $H_{\rm{o}}=100 \ \textit{h} \text{ km s}^{-1}\,\rm{Mpc}^{-1}$ with $h=0.6774$. This is consistent with \cite{Planck2016}.

\section{Methods}\label{Methods}
\subsection{Our common dark matter halo cone} \label{lightcone}

The starting point in creating our multiple versions of the mock [CII] tomographic scans was producing a common DM halo cone, that is, a catalog of DM halos positioned in a three-dimensional space with a geometry similar to that observed by a telescope. This DM cone, which extends from an imaginary observer with a reverse time evolution along the line-of-sight, has a sky coverage of a $4^{\circ} \times 4^{\circ}$ region (i.e., the envisioned size of the Prime-Cam fields) and extends from $z=0$ to $z\sim10$, that is, well within the EoR. While DM halos at $z<3.5$ are irrelevant for the present paper, they will be of utmost importance when we discuss ways of removing foreground contamination of astrophysical origin which affects the detection of the [CII] PS (Karoumpis et al. in prep.).

To construct this DM halo cone, we used the DM halo (Subfind object\footnote{Object identified by the Subfind algorithm. Many refer to the Subfind objects as subhalos, here we refer to them as halos.}) catalogs of the TNG300-1\footnote{https://www.tng-project.org/} simulation (for a detailed description of the simulation see Section \ref{DMhalos}). The DM halos were calculated using a two-step criterium: the first one consists of applying the friends-of-friends \citep[FOF,][]{Huchra} algorithm, while the second step is a selection refinement following the Subfind \citep{Rodriguez} algorithm. The FOF algorithm was applied to the full DM particle distribution in order to create groups of DM particles with an inter-particle distance lower than $ 0.2 \times$ the mean interparticle separation. Because the algorithm was applied only to the DM particles, the other types of particles (gas, stars, BHs) were assigned to the same groups as their nearest DM particle. Every group was then refined by selecting only the gravitationally bound particles in every FOF group, that are classified into halos by the Subfind algorithm, requiring each halo to contain at least 20 particles, regardless of type. We note that there is the possibility to have more than one Subfind object in a FOF group, although this is rare for the (post-)EoR reshift, for example, only $ 6 \%$ of the FOF groups of the common DM halo cone have more than one Subfind object at $ z=5.8$. In that case, the most massive object was considered as the central halo and the remaining as its satellites. Because a DM particle in the TNG300-1 simulation has a mass of $5.9\times10^{7}\,M_{\odot}$, we only considered halos with masses greater than $3\times10^{9}\,M_{\odot}$, that is, with more than 50 DM particles. We verified that above this mass range, its inferred halo mass function, agrees with the theoretical predictions from \citet{HMF} and \citet{Sheth:Tormen1999}. We note that for the redshift range of interest ($z \sim 3.5-9$) galaxies residing in halos less massive than $3\times10^{9}\,M_{\odot}$ should have a negligible contribution to the reionization as their star formation activity is suppressed by photo-heating from the intergalactic medium \citep{Thoul1996,Gnedin2000,Finlator2011,Noh2014}. 

To convert the box-shaped geometry of the TNG300-1 simulation into a cone, we first placed the observer at one of the corners of the $z=0$ cube and had them "look out" at it. We then remapped the Cartesian coordinates of the TNG300-1 halo catalog into right ascension (R.A.), declination (DEC.), and distance from the observer. We converted the distance from the observer into a cosmological redshift and added to it the peculiar velocity of DM halos along the line-of-sight to get the observed redshift. Finally, we repeated these steps, placing along the line-of-sight new data cubes so that their simulated redshift matches the cosmological redshift seen by the observer. To eliminate the occurrence of periodically repeating structures, we applied three randomization transformations on every cube: random rotation, mirroring, and translation \cite[see][]{Croton2006}.

\subsection{The connection between galaxies and their dark matter halos} \label{DMhalos}
 \begin{figure}[ht]
    \centering
        \includegraphics[width=0.53 \textwidth]{{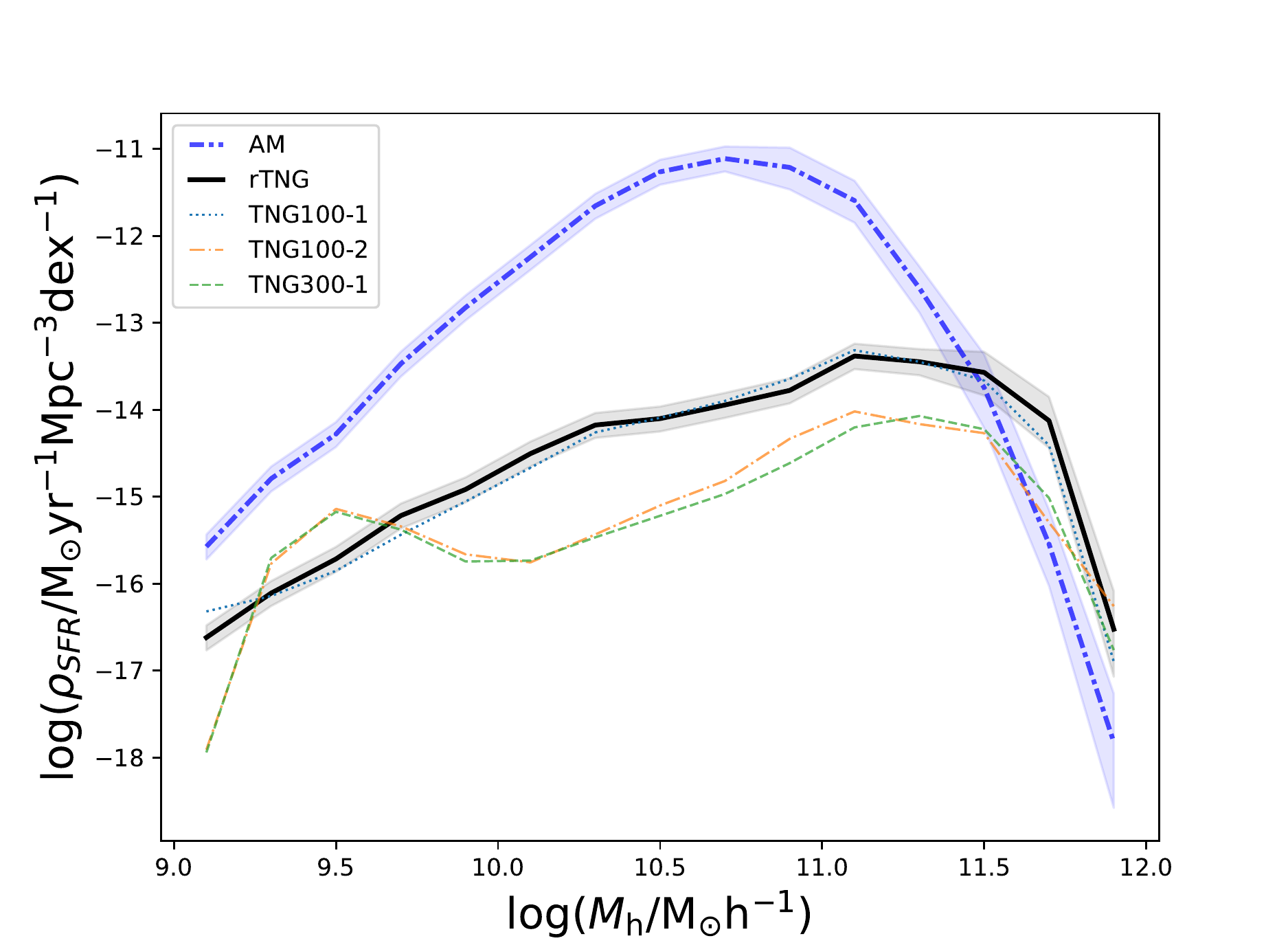}}
        \caption{Contribution of different halo mass range to the cosmic comoving star formation rate density at $z=5.8$. The dotted blue, long-dash-dotted orange, and dashed green lines are inferred from the TNG100-1, TNG100-2, and TNG300-1 simulations, respectively. The solid black line is inferred from the renormalized TNG300-1 simulation (rTNG), according to Eq.~\ref{eq:rTNG}. The dash-dotted light blue line is inferred by abundance matching (AM) our TNG300-1-based DM halo cone to the dust-corrected UV luminosity function of \citet[][see Section~\ref{AbMat}]{Bouwens:Illingworth:2015}. Gray and blue shaded areas are the $68\%$ confidence integrals for rTNG and AM results, respectively, taking into account both the effects of the Poisson error and the sample variance.}
        \label{fig:SFRD}
\end{figure}

 \begin{figure}[ht]
    \centering
        \includegraphics[width=0.53 \textwidth]{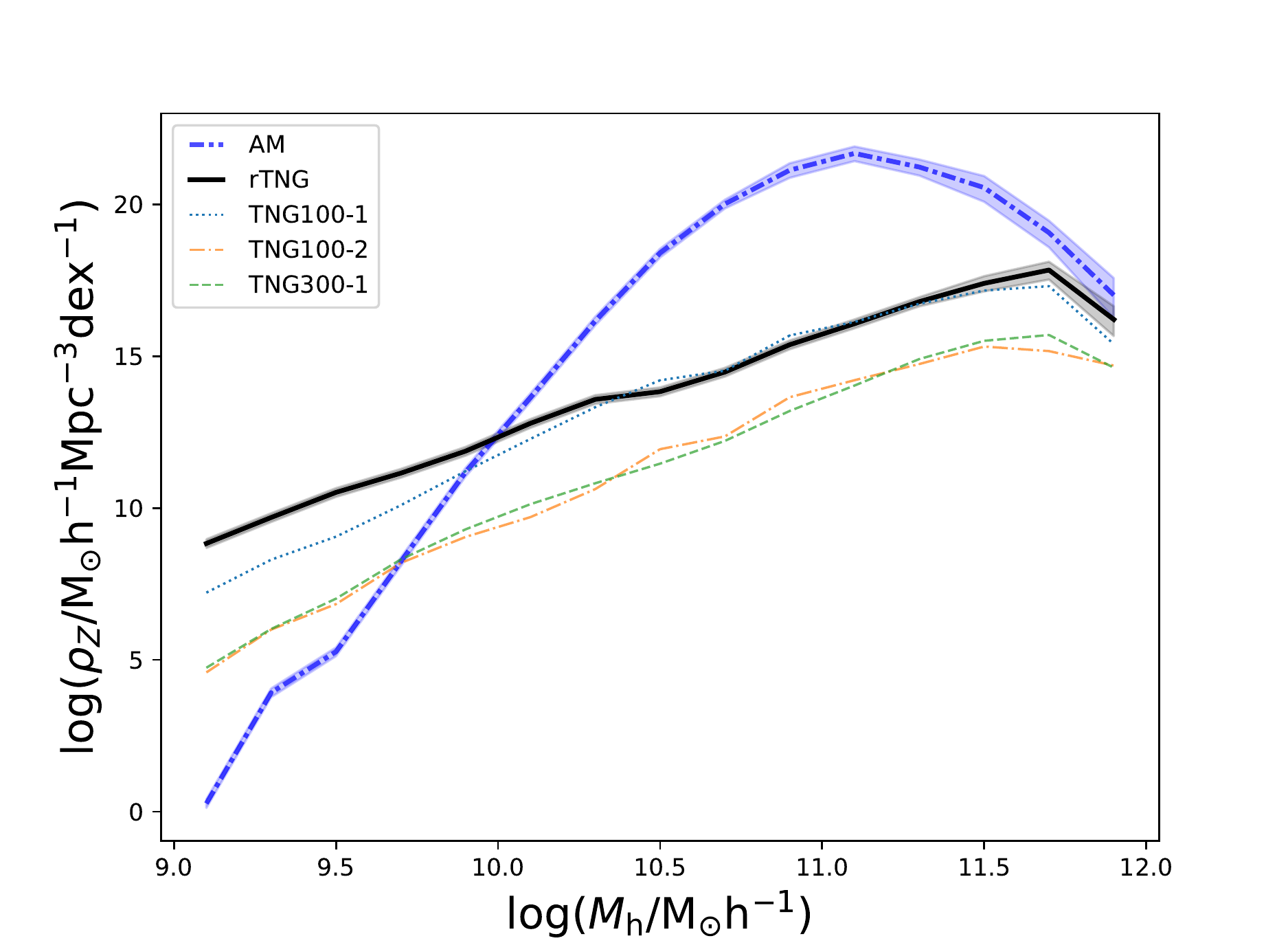}
        \caption{Contribution of different halo mass range to the cosmic comoving mass density of metals locked in stars at $z=5.8$.
        Lines and shaded areas are the same as in Fig.~\ref{fig:SFRD}. 
        }
        \label{fig:METAL}
\end{figure}

The next step in creating our [CII] tomographic scans was to populate with galaxies the DM halos of our common DM halo cone. We did so in two alternative ways. In Section~\ref{GalTNG}, we populated them with the simulated galaxies from the TNG300-1 box itself. In Section~\ref{AbMat}, we populated the halos with galaxies following an abundance matching technique assuring that the observed high-redshift dust-corrected UV luminosity function of \citet{Bouwens:Illingworth:2015} is reproduced.

\subsubsection{The galaxies of the TNG300-1} \label{GalTNG}
The most physical way to model the halo-to-galaxy relation is to use hydrodynamical simulations. Indeed, these simulations combine the gravitational evolution of the different matter constituents (i.e., DM, gas, stars, and black holes), together with the magneto-hydrodynamical behavior of the gas component.

The Illustris TNG project \citep{Pillepich:Springel:Nelson2017,Donnari2019} is an updated version of the original Illustris simulation \citep{Nelson:Pillepich2015} and is to-date the most advanced example of large hydrodynamical simulations ran in a cosmological context. Here, we made use of their largest simulation box, the TNG300-1 ($L_{\rm box}=302.6\,$Mpc, $M_{\rm DM}=5.9\times10^7\,M_{\odot}$, $M_{\rm baryon}=1.1\times10^7\,M_{\odot}$), enabling us to simulate a $4^{\circ} \times 4^{\circ}$ survey and accurately study the effect of clustering on the [CII] PS up to scales of 100\,Mpc.

The TNG300-1 halo (i.e., Subfind object; see Section 2.1) catalogs obtained from the baryonic simulation provided as well, at each time output, the properties of the galaxy\footnote{The definition of a simulated galaxy here is that of a Subfind object with stellar particles. As a result, the halos in our catalog do not host more than one galaxy, with the central halos containing the main galaxies and the rest, the satellites.} hosted by each of these DM halos. Among these properties, two of them are of particular interest for us: their SFR and metallicity from which we predict their [CII] luminosity (see Section~\ref{sfr lcii relation}).

Being the simulation with the largest box in the project, TNG300-1 is naturally not the one with the best mass and spatial resolution. Compared to the flagship TNG100-1 simulation ($L_{\rm box}=110.7\,$Mpc, $M_{\rm DM}=7.5\times10^6\,M_{\odot}$, $M_{\rm gas}=1.4\times10^6\,M_{\odot}$), it has a factor of 8 (2) lower mass (spatial) resolution. In Fig.~\ref{fig:SFRD}, we show how the cosmic SFR density inferred from these two simulations differs, especially for halos with $M_{\rm h}<10^{11}\,M_{\odot}$, where there is an order of magnitude discrepancy. The reason is that during the TNG run, there are no on-the-fly adjustments or rescaling to achieve resolution convergence \citep{Pillepich:Springel:Nelson2017}. Specifically, for star formation, the difference is related to how gas turns into stars in the Illustris TNG simulations \citep{Pillepich:Springel:Nelson2017,rTNG}. Stars form stochastically on a given timescale (i.e., $t_{\rm SFR}$) from gas cells that exceed a given density threshold (i.e., $\varrho_{\rm SFR}$). These timescales and density thresholds are the same at all resolutions, with $t_{\rm SFR }=2.2\,$Gyr and $\varrho_{\rm SFR}=0.1$ neutral hydrogen atoms per cm$^{3}$. A better spatial and mass resolution leads to the sampling of higher gas density regions, allowing more gas to fuel star formation. More SF fuel leads to a higher SFR at fixed halo mass, with increasing resolution \citep{Pillepich:Springel:Nelson2017, rTNG}. Because the metallicity is a function of stellar mass and SFR (Eq. \ref{eq:GMetal}), we notice similar discrepancies at $M_{\rm h}<10^{11}\,M_{\odot}$ between the cosmic mass density of metals (i.e., $\varrho_{\rm Z}$) predicted at $z=5.8$ by the TNG300-1 and TNG100-1 simulations (Fig.~\ref{fig:METAL}).

Thanks to the set of TNG realizations, we could quantify how the different resolutions affect the predictions of these simulations. TNG100-1 and TNG300-1 each come with a series of lower resolution realizations of the same volume, with eight and 64 times more massive DM particles (i.e., TNG100-2, TNG100-3, TNG300-2, and TNG300-3). Despite the changes in box size and initial conditions, the cosmic SFR density predicted from TNG100-2 is in very good agreement with that of TNG300-1 (Fig.~\ref{fig:SFRD}). Based on the minor influence of the different simulation volumes on the results and following \cite{rTNG}, we assumed that the outcome of the TNG100-1 simulation, which has the finest (DM and baryonic) mass resolution, is the best estimate of the galactic SFR and rescaled the SFR of the TNG300-1 galaxies on a halo-by-halo basis, as:
\begin{equation} \label{eq:rTNG}
\begin{aligned}
{\rm SFR}( M_{\rm h};\,{\rm rTNG})& =\ {\rm SFR}(M_{\rm h};\,{\rm TNG300-1})\ \times\ \\
& \frac { \overline{\rm SFR}(M_{\rm h}\in[M_{\rm h}\pm0.25{\rm dex}];\,{\rm TNG100-1}) } {\overline{\rm SFR}(M_{\rm h}\in[M_{\rm h}\pm0.25{\rm dex}];\,{\rm TNG100-2}) }\ .
\end{aligned}
\end{equation}
We applied the same halo-by-halo correction to the metallicities of each galaxies, 
\begin{equation} \label{eq:rTNGmet}
\begin{aligned}
{\rm Z}( M_{\rm h};\,{\rm rTNG})& =\ {\rm Z}(M_{\rm h};\,{\rm TNG300-1})\ \times\ \\
& \frac { \overline{\rm Z}(M_{\rm h}\in[M_{\rm h}\pm0.25{\rm dex}];\,{\rm TNG100-1}) } {\overline{\rm Z}(M_{\rm h}\in[M_{\rm h}\pm0.25{\rm dex}];\,{\rm TNG100-2}) }\ .
\end{aligned}
\end{equation}
From now on, predictions inferred from this rescaled cone catalog will be denoted "rTNG". 

Finally, as one can see in Fig.~\ref{fig:SFRD}, the curve of the cosmic SFR density inferred from TNG100-1 at $M_{\rm h}< 10^{10}\,M_{\odot}$ does not have the same shape as those inferred from the TNG100-2 and TNG300-1 simulations, which exhibit a peak at $M_{\rm{h}}= 10^{9.5} M_{\odot}$. Indeed, at $M_{\rm h}<10^{10}\,M_{\odot}$ independently of the redshift, the lower resolution of the TNG100-2 and TNG300-1 simulations drastically affect the formation and evolution of galaxies residing in such low-mass halos resulting in either an overestimation or an underestimation of their number density depending on the $M_{\rm h}$-bin. There, our simple halo-by-halo rescaling approach could not be applied, which forced us to limit this cone catalog to galaxies hosted by $M_{\rm h}>10^{10}\,M_{\odot}$ halos. The importance for the [CII] PS of (post-)EoR galaxies located in DM halos with $3\times10^{9}< M_{\rm h}/M_{\odot} <10^{10}$ was thus only assessed using the alternative method of abundance matching.

\subsubsection{Abundance matching} \label{AbMat}

Abundance matching (AM) is a method based on the simple hypothesis that the most massive galaxies occupy the most massive halos. Starting from the observed stellar mass function at a given redshift, one creates a mock population of galaxies and then matches them accordingly to the halos.
The DM mass is, however, not the only halo property that can be matched to the observed stellar mass function. According to the latest studies of the halo-to-galaxy relation \citep{Wechsler:Tinker2018}, it is not even the optimal choice: using mass as the matching quantity neglects the fact that when a halo enters the gravity field of a larger neighboring halo, it is affected by intense tidal stripping; matching the mass of this halo after the stripping would hence result into incorrect galaxy properties. Therefore, we chose to perform our abundance matching technique using a halo property less influenced by the tidal stripping than its mass. Following \cite{Kravtsov2004} and \cite{Bethermin2017}, we used the maximum value of the spherically-averaged rotation curve of the halo, that is, $V_{\rm max}$.

Unfortunately, there exists to date no observational constraints on the galaxy stellar mass function at the EoR. Fortunately, at such high redshift, performing AM using a SFR function well constrained by \textit{HST} observations is to first order as appropriate as using a stellar mass function \citep{Yue:Ferrara2015,Yue:Ferrara2019}. Indeed, at $z \gtrapprox 4$, the probability for a galaxy hosted on a $\rm M_{\rm h} >10^{10}\,M_{\odot}$ halo to be non-SF is close to zero \citep{Bethermin2017} and there exists a tight correlation between the stellar mass and the SFR of SF galaxies at all redshifts probed to-date, the so-called main sequence of SF galaxies \citep[e.g.,][]{MainSeq}. We also note that performing the AM using the SFR function instead of the stellar mass function has the advantage to reduce the number of steps from AM to $L_{\rm [CII]}$, as the [CII] luminosity of high-redshift galaxies is expected to scale with their SFRs \citep[e.g.,][]{Olsen2015,Vallini2015,Lagache2018,ALPINE}. We started, therefore, from the observed UV luminosity function, which is described by a Schechter function \citep{Schechter}, written in terms of magnitude:
\begin{equation}
  \frac{dn}{dM_{\rm{UV}}} =0.4 \ \ln(10) \ \phi_{*} \ x^{(1+ \alpha)} \ e^{-x}\ , 
\end{equation}
where $x=10^{0.4 \ (M^{*}_{\rm{UV}}-M_{\rm UV})}$, with $M_{\rm{UV}}$ being the dust-attenuated absolute AB magnitude, $\alpha$ is the faint-end slope parameter, $\phi_{*}$ is the characteristic number of galaxies per comoving volume, and $M^{*}_{\rm UV}$ is the characteristic absolute magnitude, at which the luminosity function exhibits a rapid change in its slope \citep{Schechter}. According to \cite{Bouwens:Illingworth:2015}, in the redshift range 4 to 9:
\begin{eqnarray}
\begin{aligned}
 M^{*}_{\rm{UV}}&=-20.96+0.01 \ (z-6)\ , \\
 \phi _{*}&=0.46 \times 10^{-3\,-0.27 \ (z-6)},\\ 
 \alpha&=-1.87 - 0.10 \ (z-6)\ .
\end{aligned}
\end{eqnarray}
In order to derive the SFR of our mock galaxy population, we needed to take into account the dust attenuation and calculate the dust-corrected UV luminosity function, which subsequently could be converted into a SFR function. Following \cite{Yue:Ferrara2019}, we used the coefficients of \cite{Koprowski}, where the dust-corrected absolute magnitude is
\begin{equation}
M^{'}_{\rm{UV}}=M_{\rm{UV}}-A_{1600}\ , 
\end{equation}
where 
\begin{equation}
A_{1600}=4.85+2.10 \ \beta(\geq 0)\ ,
\end{equation}
is the dust attenuation at 1600\ ,{\AA} and $\beta$ is the measured UV spectral slope, that is, 
\begin{equation}
f_{\lambda} \varpropto \lambda ^{\beta}\ .
\end{equation}
The spectral slope $\beta$ depends on $M_{\rm{UV}}$ and was fitted by 
\begin{equation}
\beta = \beta_{-19.5}+ \frac{d \beta}{d M_{\rm{UV}}} (M_{\rm{UV}} + 19.5)\ .
\end{equation}
From \cite{Bouwens:Illingworth:2015} we have:
\begin{eqnarray}
\begin{aligned}
& \beta _{-19.5}= -1.97 - 0.06 \ (z-6)\ ,  \\
& \frac{d \beta}{d M_{\rm{UV}}}=-0.18 - 0.03 \ (z-6)\ .
\end{aligned}
\end{eqnarray}

The dust-corrected UV luminosity function was then related to the measured UV luminosity function via 
\begin{equation}
    \frac{dn^{'}}{dM^{'}_{\rm{UV}}}(M^{'}_{\rm{UV}},z)= \frac{dn}{dM_{\rm{UV}}}(M_{\rm{UV}},z)\ .
\end{equation}

We assumed that the dust-corrected UV luminosity is linked to the DM mass halo function via a monotonic $M_{\rm h}-M_{\rm UV}^{'}$ relation with a 0.2\,dex log-normal scatter \citep{Corasaniti2017}. In order to match our dust-corrected luminosity function to the halo mass function, we had to assume a monotonic function without scatter: this way, there is an exact, "direct" match of DM halos ordered by their mass and galaxies ordered by their UV magnitude. To get to this direct luminosity function, we re-wrote the dust corrected UV luminosity function as: 

\begin{equation}
    \frac{dn^{'}}{dM^{'}_{\rm{UV}}}= \left( \frac{dn^{'}}{dM^{'}_{\rm{UV}}} \right ) _{\rm{direct}} * P_{\rm{s}}\ ,
\end{equation}

where "$*$" denotes the convolution operation and $P_{\rm{s}}$ describes the log-normal scatter of the $M_{\rm h}-M_{\rm UV}^{'}$ relation. $ M_{\rm h}-M'_{\rm UV,\ direct}$ is a monotonic relation without scatter, suitable for the AM method, that is,

\begin{equation}
    \left( \frac{dn^{'}}{dM^{'}_{\rm{UV}}} \right ) _{\rm{direct}}= \frac{dN}{d\log M_{\rm{h}}} \frac{d \log M_{\rm{h}}}{dM^{'}_{\rm{UV}}}\ ,
\end{equation}
where $N$ is the number of DM halos. To perform the abundance matching of our dust-obscured UV luminosity function to the DM halo mass function via this direct dust-obscured UV luminosity function technique, we used the code of Yao-Yuan Mao\footnote{https://bitbucket.org/yymao/abundancematching/src/master/} that provides a python wrapper around the deconvolution kernel described in \citet{Behroozi:Conroy:2010}. In this way, we obtained for each DM halo within our cone, the direct dust-corrected UV absolute magnitude ($M_{\rm UV}^{'{\rm ,\ direct}}$) and the "observed" dust-correct UV absolute magnitude (i.e., $M_{\rm UV}^{'}$) of its embedded galaxy. We should stress that these direct magnitudes were only used to perform the abundance matching, while the SFR of galaxies residing in these DM halos were computed from their observed magnitude. To do so, we converted their dust-corrected UV absolute magnitude, 
into SFR following \citet{Kennicutt98}, that is,
\begin{equation}
\text{SFR}=  \frac{ 4 \pi \times 3.08\times10^{19} \times 10^{ -0.4 \ (M_{\rm UV}^{'} + 48.6)}}{l_{\rm UV}}\ ,
\end{equation}

where $l_{\rm UV}=8.9 \times 10^{27}\,$erg\,s$^{-1}$Hz$^{-1}\,(M_{\odot}/$yr$)^{-1}$ \citep[][assuming a metallicity of $0.1\,Z_{\odot}$, a stellar age of 10\% the Hubble time, and a Salpeter initial mass function between $0.1-100\,M_{\odot}$]{Yue:Ferrara2015}.

 \begin{figure}[t]
    \centering
        \includegraphics[width=0.53 \textwidth]{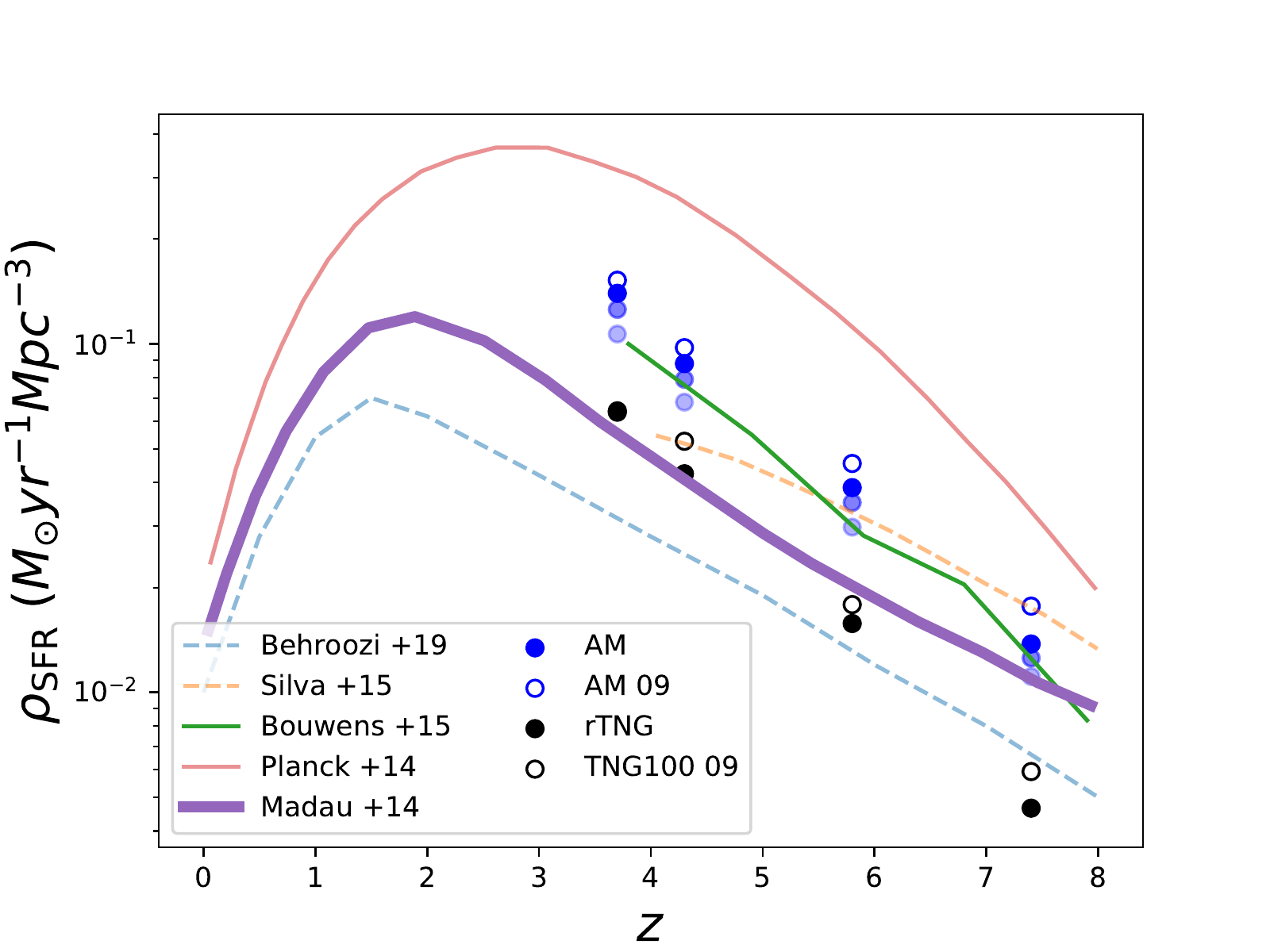}
        \caption{Redshift evolution of the cosmic SFRD as inferred from our rTNG (black dots) and AM (dark blue dots) models. Our AM results are also presented for a slightly different dust correction similar to that used in \citet[][blue dots]{Bouwens:Illingworth:2015} and for a cone catalog restricted to galaxies not brighter than the brightest SF galaxies observed in the pencil-beam survey of \citet[][light blue dots]{Bouwens:Illingworth:2015}. The empty blue and black circles is our AM and TNG100 prediction for the lower halo mass limit of $M_{\rm h}=3 \times 10^9 M_{\odot}$. The solid green, red, and purple lines present observational constraints from \citet[][]{Bouwens:Illingworth:2015}, \citet[][]{P14}, and \citet[][]{Madau:Dickinson2014}, respectively. The dashed blue line is the UNIVERSE Machine prediction \citep{Behroozi} and the dashed orange line is the SFRD that comes from analytically integrating the $M_{\rm h}$-to-SFR relation of \citet[][]{Silva2015} over all halo masses.}
        \label{SFRDOTHERS}
\end{figure}

One of the $L_{\rm [CII]}-$SFR coupling relations used \citep[][introduced in Section \ref{sfr lcii relation}]{Vallini2015}, also depends on the metallicity of our galaxies. Those metallicities were inferred using the fundamental metallicity relation \citep{Mannucci},
\begin{eqnarray}
\begin{aligned}
\log (Z) =&\ 0.21+0.37 \log \left(M_{\rm \ast}/10^{10}\right)-0.14 \ \log (\mathrm{SFR})  \\
& -0.19 \ \log ^{2}\left(M_{\rm \ast}/10^{10}\right)-0.054 \log ^{2}(\mathrm{SFR}) \\
& +0.12 \ \log \left(M_{\rm \ast}/10^{10}\right) \ \log (\mathrm{SFR}),
\end{aligned} \label{eq:GMetal}
\end{eqnarray}
where the stellar mass, $M_{\rm \ast}$, of our AM galaxies was obtained from the latest measurement of the mass-to-UV-light ratio \citep{Duncan},
\begin{equation}
\log \left(M_{*} \right)=-1.69-0.54 \ M_{\mathrm{UV}}\ .
\end{equation}
The fact that the DM halos were connected to the [CII] emission according to the above observational relations means that the only mass resolution that limits the AM method is that of the DM particle. With this method, we could thus investigate the influence of galaxies formed in low mass halos --that is, $3\times10^{9}<M_{\rm h}/M_{\odot}<10^{10}$-- not accounted for in our rTNG simulation. We did so by applying the AM twice: assuming a DM halo mass lower limit of $10^{10} M_{\odot}$ (i.e., matching the limit of our rTNG catalog) and considering an even lower limit of $3\times10^9\, M_{\odot}$.\\

In Fig.~\ref{SFRDOTHERS}, we compare the redshift evolution of the cosmic SFR densities (SFRD) as predicted by our rTNG and AM models to observational constraints and simulation predictions from the literature. Our rTNG predictions are in good agreement with measurements from \citet{Madau:Dickinson2014}, which at these redshifts are mostly based on the dust-corrected UV luminosity functions of \cite{Bouwens12B, Bouwens12A}. On the contrary, our AM predictions lie significantly above $\times(2-3$) these observations, although with values not as high as constraints from the \citet{P14}. This was to be expected because the SFRD inferred in \citet{Bouwens:Illingworth:2015} and from which our AM model is based also lies above the measurements of \citet{Madau:Dickinson2014}. We note, however, a $20-30\%$ disagreement between our AM-based SFRD and those inferred in \citet{Bouwens:Illingworth:2015}. Part of this disagreement comes from adopting here different dust correction and $l_{UV}$ values than in \citet[][see blue dots in Fig.~\ref{SFRDOTHERS}]{Bouwens:Illingworth:2015}; furthermore our large simulated volume contains galaxies with higher SFRs than the brightest SF galaxies observed in \citet[][see the light blue dots in Fig.~\ref{SFRDOTHERS}]{Bouwens:Illingworth:2015}. Last, our AM-based SFRD increases by $10-30\%$ when accounting for the contribution of galaxies residing in $3\times10^{9}<M_{\rm h}/M_{\odot}<10^{10}$ halos and not accounted for in our rTNG simulation (Fig.~\ref{SFRDOTHERS}). For comparison we also plot the SFRD of the TNG100 snapshots, applying also the lower ($ \rm 3\times10^{9}<M_{\rm h}/M_{\odot}$) DM halo mass limit, getting $\rm 0-10\%$ higher values than the SFRD of rTNG. 

Finally, to test for any potential influence of baryonic substructures (i.e., massive DM halos hosting more than one massive galaxy) on our AM result, we reapplied our method but this time by matching the UV luminosity function with the rTNG galaxies, based on their stellar masses, instead of the rTNG DM halos (Fig.~\ref{fig:SFRD} and Fig.~\ref{fig:METAL}). As it can be noticed, there is no significant deviation from the original AM result for $ M_{\rm h}>10^{10}\, M_{\odot}$, suggesting that the number of halos hosting more than one galaxy is insignificant at our redshifts and halo mass bins of interest. The two results differ only at $ M_{\rm h}<10^{10}\, M_{\odot}$ where the low mass resolution of the TNG300-1 simulation significantly affects its baryonic matter predictions (see Section \ref{GalTNG}), making the result of AM with DM halos more reliable.

\subsection{The [CII] emission of galaxies at the (post-)EoR} \label{sfr lcii relation}
 \begin{figure}[t]
    \centering
        \includegraphics[width=0.53\textwidth]{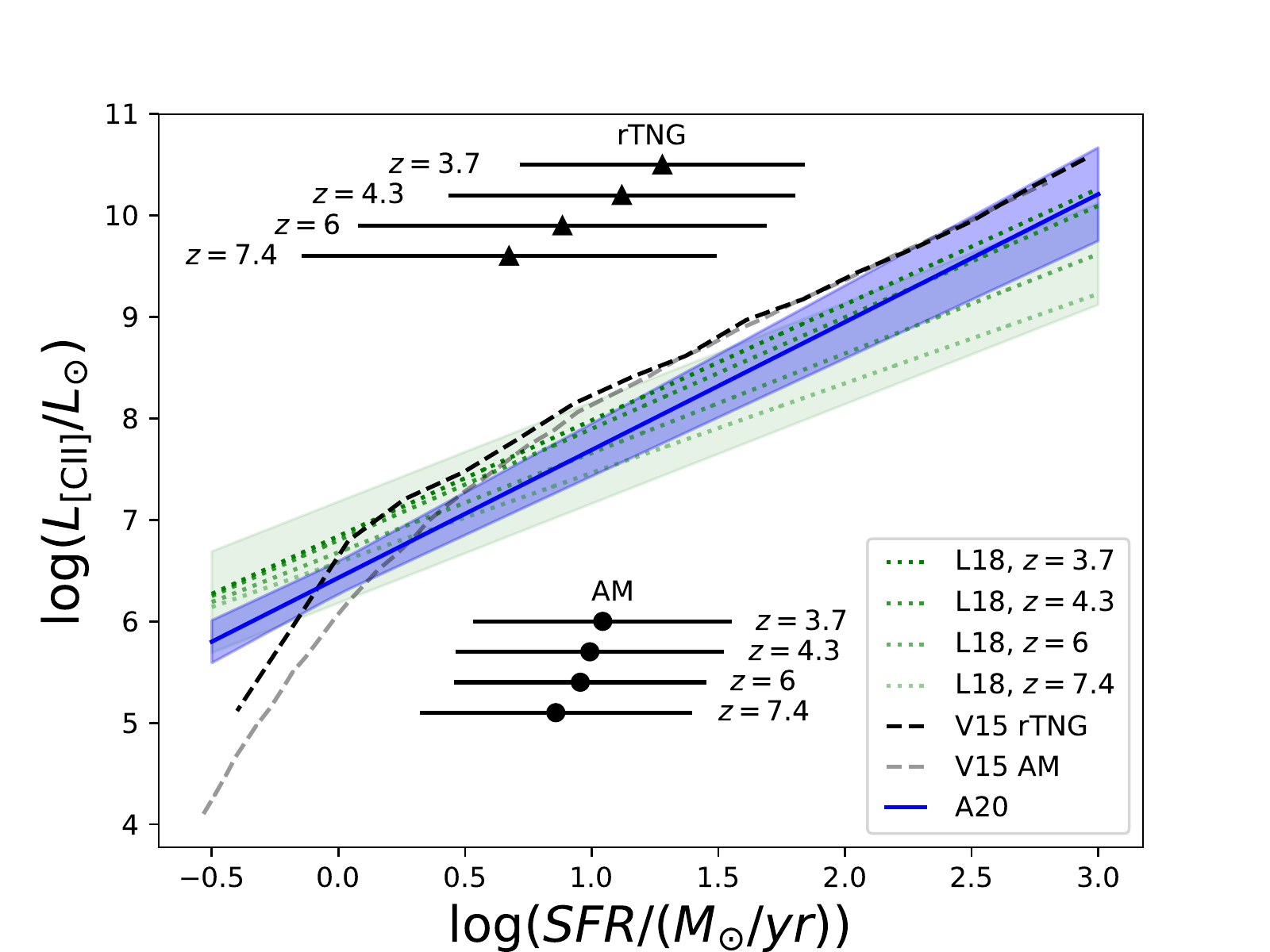}
        \caption{${\rm SFR}-L_{\rm [CII]}$ relation applied to our cone, as predicted in L18 at $z\sim 3.7, 4.3, 5.8$, and $7.4$ (green dotted lines) and observed in A20 at $4.4<z<9.1$ (blue line) along with its $1\sigma$ scatter (blue shaded region). For clarity, the redshift-independent $1\sigma$ scatter of 0.5\,dex inferred in L18 is only shown around their $z=5.8$ mean relation (green shaded region). The mean metallicity-dependent ${\rm SFR}-L_{\rm [CII]}$ relation of V15 applied to our rTNG and AM models are shown by the dark and light gray dashed lines, respectively. The horizontal black lines represent the SFR ranges containing $25\%-75\%$ of the cumulative cosmic SFRD at $z\sim 3.7, 4.3, 5.8,$ and $7.4$ in our rTNG (triangles) and AM (circles) models. These ranges highlight the SFRs of galaxies that contribute the most to the cosmic SFRD at these redshifts.}
        \label{fig:LCII}
\end{figure}
The correlation between the [CII] luminosity and the SFR of galaxies results from the balance between the stellar feedback heating up the gas and the ability of [CII] to cool it by radiating energy away. Despite the simplicity of this premise, modeling the exact physics of the [CII] line emission is complex, as it originates from various phases of the interstellar medium (ISM). Those include photo-dissociation regions (PDRs), the warm ionized medium (WIM), and the warm and cold neutral medium \citep[WNM,CNM;][]{Croxall2017,Madden,Kaufman,Carpio,Cormier,Appleton,Velusamy2014,Pineda,Croxall2017}. 
In spite of these intricacies, a tight relation between SFR and $L_{\rm [CII]}$ has been reported in local galaxies \citep{DeLooze,Herrera}.
This relation seems to hold at $z>4$, albeit with an increasing scatter \citep{Carniani,Fujimoto,ALPINE,FujimotoALPINE}. This scatter is, however, not surprising and actually predicted by hydrodynamical simulations and semi-analytical models \citep{Vallini2015,Pallottini2015,Pallottini2017,Pallottini:Ferrara:Bovino2017,Olsen2017,Katz,Lagache2018}. It seems to result from the interplay of different factors such as variation in metallicities, gas mass, and interstellar radiation fields of the galaxies during the (post-)EoR. 
 
In this paper, we used three different scaling relations to study their influences on our [CII] PS forecasts:  
\begin{enumerate}
\item \citet[][hereafter L18]{Lagache2018} assume that the bulk of the [CII] luminosity of high-redshift galaxies comes from their PDRs. They use a semi-analytic model of galaxy formation combined with the photo-ionization code \texttt{Cloudy} \citep{Ferland13,Ferland} to calculate the luminosity of 28,000 mock galaxies at $z>4$. They find that the [CII] luminosity is a function of the SFR and the redshift of a galaxy,
\begin{equation}
\log\left (\frac{L_{\rm{[CII]}}}{L_{\odot}}\right )=(1.4-0.07 \ z) \ \log\left (\frac{{\rm SFR}}{M_{\odot}\,{\rm yr}^{-1}}\right )+7.1-0.07 \ z \ ,
\end{equation}
with a $\sim 0.5\,$dex scatter.

\item \citet[][hereafter V15]{Vallini2015} model a two-phase ISM consisting of PDRs and a CNM. They combine a radiative transfer hydrodynamical simulation of a $z=6.6$ galaxy located in a $M_{\rm h}=1.7 \times 10^{11} \, M_{\odot}$ halo with a subgrid ISM model and the PDR-code \(\texttt{UCL\_PDR}\) \citep{Bell2005,Bell2007,Bayet2009}. Running their subgrid model for a range of SFR values, $\text{SFR} =[0.1-100 \, M_{\odot} \, yr^{-1}]$, they find that the [CII] luminosity depends not only on the SFR but also on the metal content of galaxies, 
\begin{eqnarray}
  \log\left (\frac{L_{\rm{[CII]}}}{L_{\odot}}\right )=7.0+1.2 \ \log\left(\frac{\rm{SFR}}{M_{\odot}\rm{yr}^{-1}}\right )+0.021\ \log\left(\frac{Z_{g}}{Z_{\odot}}\right ) \nonumber \\
+0.012 \ \log\left (\frac{\rm{SFR}}{M_{\odot}\rm{yr}^{-1}}\right ) \ \log\left(\frac{Z_{g}}{Z_{\odot}}\right )-0.74 \ \log^{2}\left (\frac{Z_{g}}{Z_{\odot}}\right ). 
\end{eqnarray}
There is no explicit redshift evolution and scatter predictions for this relation as it is derived from a $z=6.6$ simulation. A scatter of $\sim0.2$\,dex and $\sim0.1$\,dex is, however, implicitly introduced by the distribution of metallicity at a given SFR in our rTNG and AM simulations. A redshift evolution is also implicitly introduced by the slight increase of the mean metallicity of galaxies from $z\sim7.4$ to $z\sim3.7$, but this evolution is mostly insignificant for these mean relations. Finally, we note that the mean relations inferred from the rTNG and AM models significantly differ at $\text{SFR}<1\,M_{\odot}\, \rm{yr}^{-1}$, that is, SFRs where in the AM model galaxies have significantly lower metallicities and therefore lower [CII] luminosities than in the rTNG simulation. 

\item \citet[][hereafter A20]{ALPINE} combine 75 [CII] robust detections and 43 upper limits, obtained by the ALMA Large Program to INvestigate C + (ALPINE) survey \citep{LeFevreALPINE}, with 36 earlier [CII] observations. They gather a sample of 154 main-sequence galaxies located between $4.4<z<9.1$. According to their Bayesian fit, the [CII] luminosity of a galaxy and its observational scatter are correlated to its SFR as:
\begin{equation}
\log\left (\frac{L_{\rm{[CII]}}}{L_{\odot}}\right )=(6.43 \pm 0.16)+(1.26 \pm 0.10) \ \log\left (\frac{\rm{SFR}}{M_{\odot}\rm{yr}^{-1}}\right )\ . \end{equation}
\end{enumerate}
 
While these three relations are overall in good agreement, their associated scatter considerably differs (Fig.~\ref{fig:LCII}). On the one hand, one should note that the scatter of A20 includes observational uncertainties and possible selection biases. Indeed, although the number of [CII] detections of high-redshift galaxies is growing, it is still not large enough for a detailed statistical analysis. On the other hand, calculating the intrinsic scatter with simulations is also challenging: hydrodynamical simulations of high-redshift galaxies still suffer from the small number of simulated objects and the limits imposed by the mass resolution on the modeling of the ISM. Therefore, one should be aware that scatter either reported from the observations or predicted by the simulations is still uncertain. We examine in Sect.~\ref{mean intensity} and Sect.~\ref{PS} the influence of this scatter on the predicted mean [CII] line intensities and PS, respectively. 

\subsection{[CII] tomographic scans of the (post-)EoR}
From the cone catalogs generated in the previous sections, we created our mock three-dimensional tomographic scans, that is, data cubes in which each slice corresponds to a $4^{\circ} \times 4^{\circ}$ region of the sky and contains the cumulative [CII] emission of galaxies within a particular redshift range (equivalently frequency range). The properties of these three-dimensional tomographic scans--that is, frequency and spatial resolutions--were tailored to the specifications of the two spectrally/spatially multiplexing Fabry-Perrot interferometers that will be placed in front of two of the Prime-Cam modules \citep{vavagiakis/etal:2018}. This EoR spectrograph will observe the sky using four spectral windows of 40\,GHz bandwidth each, that is, probing the [CII] line emitted at ($ \rm z=[6.76-8.27], [5.34-6.31], [4.14-4.76], [3.42-3.87]$), centered at $\nu_{\rm [CII]}/(1+7.45) = 225\,$GHz, $\nu_{\rm [CII]}/(1+5.79)=280\,$GHz, $\nu_{\rm [CII]}/(1+4.43)=350\,$GHz, and $\nu_{\rm [CII]}/(1+3.64)= 410\,$GHz; with spectral resolutions of 2.1, 2.7, 3.6, and 4.4\,GHz; and beam full width at half maximum (FWHM) of 0.88, 0.77, 0.65 and 0.62\,arcmin, respectively.
\begin{figure}[t]
    \centering
        \includegraphics[width=0.53\textwidth]{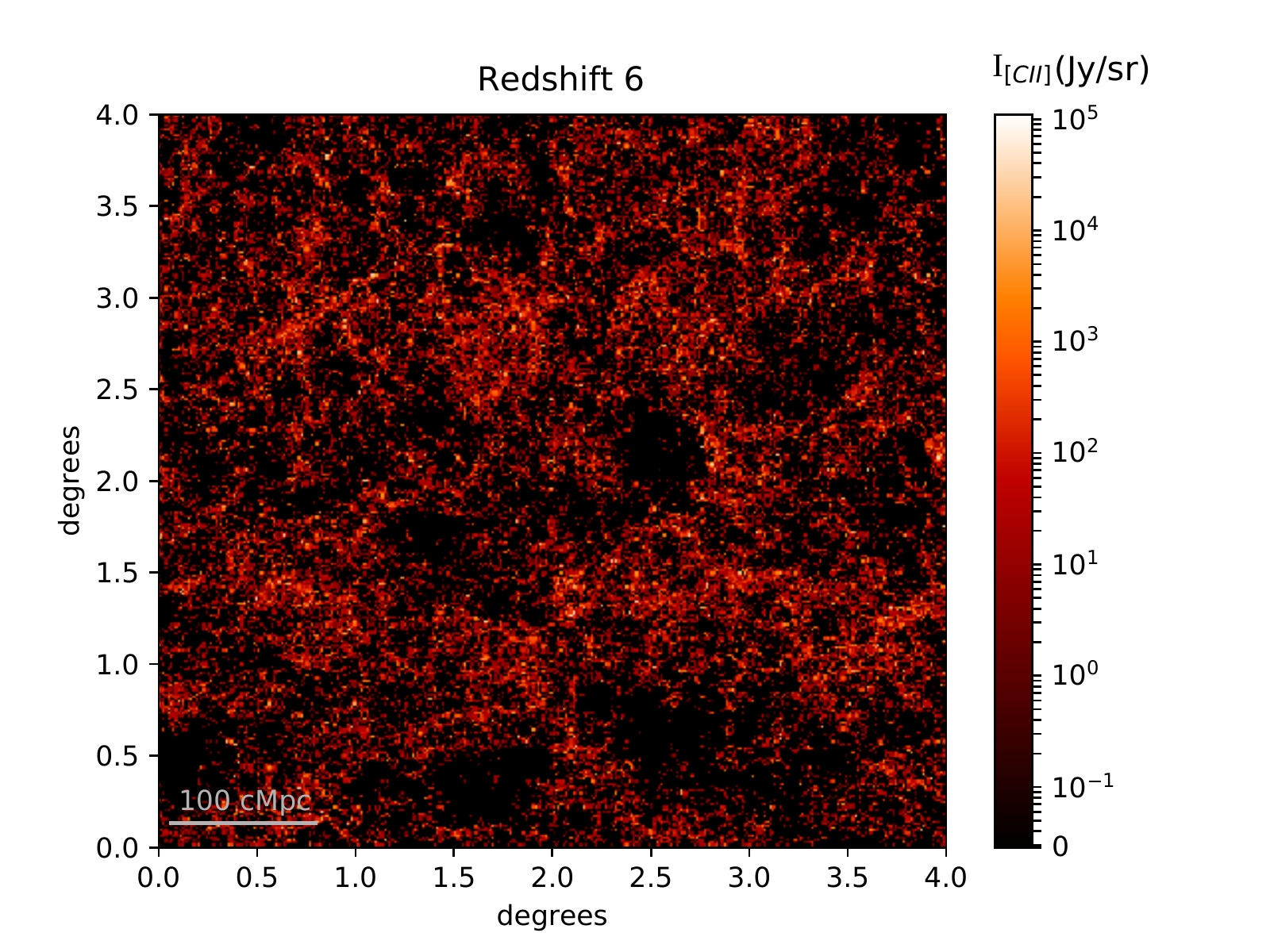}
        \caption{[CII] line intensity map at $z=5.8$ ($\nu_{\rm obs}=280\,$GHz) as predicted from our rTNG cone catalog using the SFR-$L_{\rm [CII]}$ scaling relation of \cite{Vallini2015}.}
        \label{fig:map}
\end{figure}
For each of our cone catalogs, we thus generated four tomographic scans according to the beam size and spectral resolution of these four spectral windows. As a result, our 225, 280, 350, and 410\,GHz tomographic scans include ($253 \times 253 \times 19$), ($313 \times 313 \times 15$), ($369 \times 369 \times 11$), and ($387 \times 387 \times 9$) voxels, respectively. The measured [CII] intensity in one of these voxels with observed central frequency, $\nu_{0}$, is:
\begin{equation} \label{eq:map}
I_{\rm [CII]}=\frac{1}{( \Delta \theta_{\rm b})^2} \sum \frac{1}{\Delta \nu_0} \frac{L_{\rm [CII]}^j}{4 \pi \ r ^2_j(1+z_j)^2}\ ,
\end{equation}
where $\Delta \theta _{\rm b}$ is the angular size of the voxel, $\Delta \nu_{0}$ is its bandwidth, and $r_{j}$ is the comoving distance of the $j$-th galaxy which resides at a redshift $z_{j}$ and have a [CII] luminosity $L_{\rm [CII]}^{j}$. We summed over every galaxy in the line-of-sight of our voxel that has a redshift yielding [CII] observed-frame frequency emission within the frequency range of our voxel, that is, 
\begin{equation}
    \frac{\nu_{\rm [CII]}}{\nu_{0} + \Delta \nu_{0}/2} - 1 \leq z_{j} \leq \frac{\nu _{\rm [CII]}}{\nu_{0}- \Delta \nu_{0}/2} -1\ .
\end{equation}

Fig.~\ref{fig:map} presents one "slice" of our [CII] tomography, corresponding to the [CII] line intensity map at $z=5.8$ ($\nu_{\rm obs}=280\,$GHz) as predicted from our rTNG cone catalog using the SFR-$L_{\rm [CII]}$ scaling relation of V15. This map provides a visual intuition for the dimensions of the survey and the cosmological structures enclosed in it.

\section{Results} \label{Results}
\subsection{ The mean [CII] line intensity} 
\label{mean intensity}
 \begin{figure}[t]
    \centering
        \includegraphics[width=0.53 \textwidth]{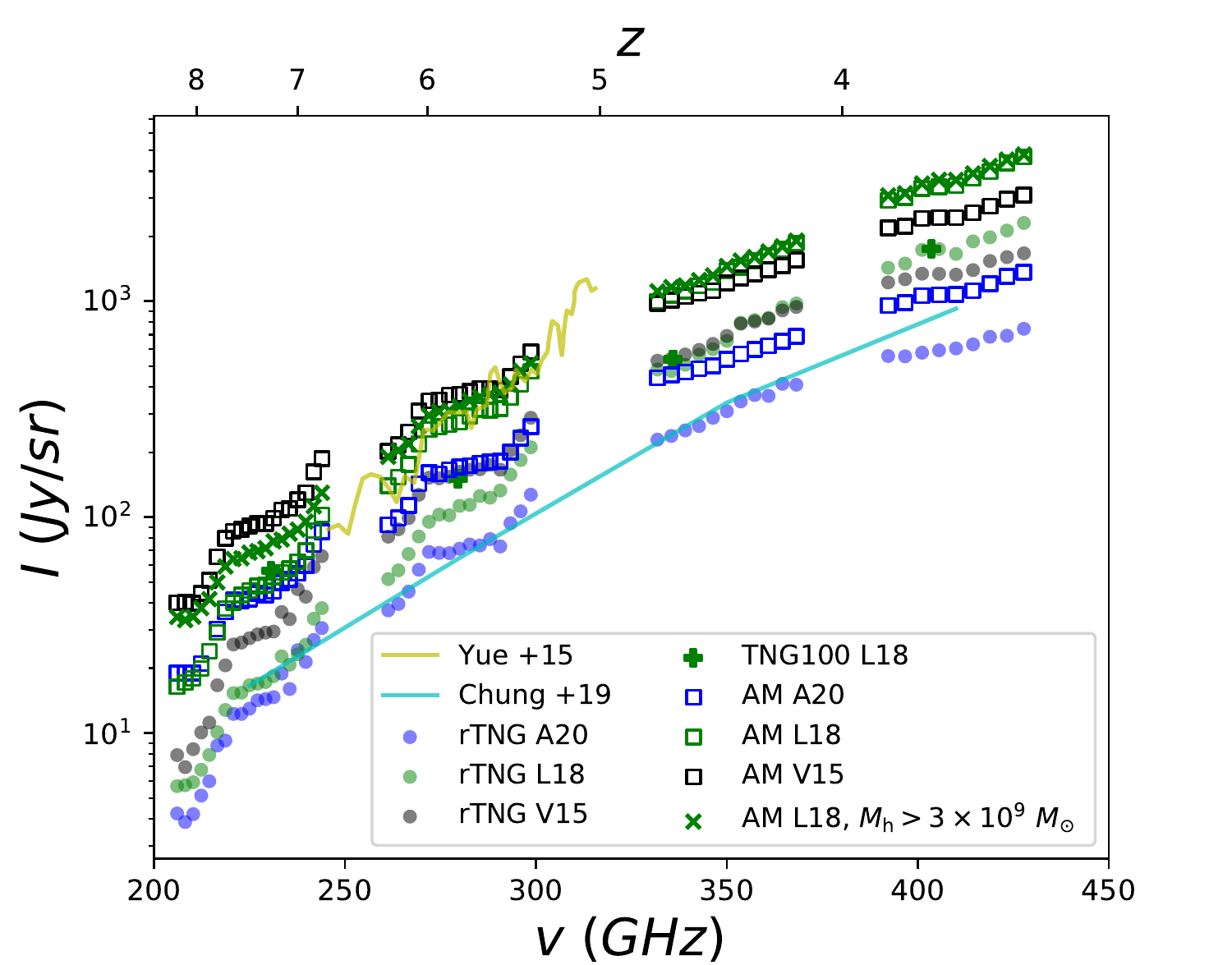}
        \caption{Mean [CII] line intensity as a function of the observed frequency (equivalently, emitted redshift). Predictions from our rTNG and AM models are shown by dots and open squares, respectively. Symbols are color-coded according to the used SFR-to-$L_{\rm{[CII]}}$ relation, blue symbols for A20, green symbols for L18, and black symbols for V15. L18 is also plotted including the contribution of $3 \times 10^{9}-10^{10} \, M_{\odot}$ halos for our AM model (green x-shaped points) and the TNG100-1 snapshots (green plus-shaped points; see text for more details). Predictions from \citet[][yellow line]{Yue:Ferrara2015} and \citet[][turquoise line]{Chung:Viero2018} are also shown for comparison.}
        \label{fig:PIX}
\end{figure}

In this section, we explore the effect of our different modeling approaches on the predicted mean [CII] line intensities emitted by (post-)EoR galaxies as a function of redshifts (equivalently observed frequencies).

The mean [CII] intensity, $\bar{I}_{\rm[CII]}$, estimated for all frequency channels of all our mock three-dimensional tomographic scans is shown in Fig.~\ref{fig:PIX}. All models predict a significant drop in $\bar{I}_{\rm[CII]}$ as we move to lower observing frequencies, equivalently to higher redshifts. This drop results naturally from (i) the cosmological dimming of the flux density of higher-redshift galaxies and (ii) the decline of the cosmic SFRD from $z\sim3$ to $z\sim8$ (Fig.~\ref{SFRDOTHERS}). Despite all models following this general redshift trend, there is up to an order of magnitude offset between their predictions. Firstly, forecasts based on the same SFR-to-$L_{\rm{[CII]}}$ relation but different halo-to-galaxy SFR relation differs by a factor two at $z=3.7, 4.3$, and 5.8 and a factor of three at $z=7.4$, with the AM model yielding systematically higher mean [CII] line intensities. Secondly, forecasts based on the same halo-to-galaxy SFR relation but different SFR-to-$L_{\rm{[CII]}}$ relations also exhibit differences: at all redshifts, V15 yields a factor of two higher mean [CII] line intensities than A20; L18, the only relation with a significant redshift evolution, yields mean [CII] line intensities that are similar to those based on V15 at $z=3.7$ and $4.3$, and lie between V15 and A20 at $z=5.8$, and in agreement with A20 at $z=7$. 

To track the origin of these differences to some specific parameters in our models, we study the analytical form of $\bar{I}_{\rm{[CII]}}$, which is given in unit of Jy/sr by,
\begin{equation}
\bar{I}_{\rm{[CII]}} = \int \frac{L_{\rm{[CII]}}}{4 \pi D_{L}^{2}}  \,y_{\rm{[CII]}}\,D_{A}^{2}\,\frac{dn}{d \log L_{\mathrm{[CII]}}}\,d \log L_{\mathrm{[CII]}}\ ,
\end{equation}
where $y_{\rm{[CII]}}=\lambda_{\mathrm{[CII]}, \mathrm{rest}}(1+z)^{2} / H(z)$ is the derivative of the comoving radial distance with respect to the observed frequency, and $D_{\rm{A}}$ and $D_{\rm{L}}$ are the comoving angular and luminosity distances \citep[e.g.,][]{Uzgil2014}. Then, assuming a generalized form of the SFR-$L_{\rm{[CII]}}$ relation,
\begin{equation} \label{generalSFRL}
\log \left ( \frac{L_{\rm{[CII]}}}{L_{\odot}}\right )=A \ \log(\text{SFR})+B+ \sigma _{\rm{L}}\ ,
\end{equation}
we can write,
\begin{equation} \label{ISFR}
\bar{I}_{\rm{[CII]}} = 10^{B} \int \frac{dn}{d \log(\text{SFR})} 10^{( \sigma _{\rm{L}} ^{2}/2)} \frac{\text{SFR}^{A}}{4 \pi D_{L}^{2}} \ y_{\rm{[CII]}} \ D_{A}^{2} \ d \log \text{(SFR)}\ ,
\end{equation}
where the factor $10^{( \sigma _{\rm{L}} ^{2}/2)}$ comes from the log-normal form of the SFR-to-$L_{\rm{[CII]}}$ relation and which implies that,
\begin{equation}
\bar{L}_{\rm{[CII]}}= \text{med}(L_{\rm{[CII]}}) \times 10^{( \sigma_{\rm{L}}^{2}/2)}\ ,
\end{equation}
where $\text{med}(L_{\rm{[CII]}})$ is the median value of $L_{\rm [CII]}$. Given that $A\sim1$, a valid assumption within the SFR ranges that contribute the most to the SFRD (see horizontal lines in Fig.~\ref{fig:LCII}), it comes from Eq.~\ref{ISFR} that the mean [CII] intensity is proportional to $\int \frac{dn}{d \log(\text{SFR})} \text{SFR} \ d \log \text{(SFR)}$, that is, the cosmic SFRD. This explains the discrepancies between the rTNG and AM models that share the same SFR-to-$L_{\rm [CII]}$ relation: the factor of two in $\bar{I}_{\rm[CII]}$ at $z=3.7$, $4.3$, and $5.8$ and the factor of three at $z=7.4$ can be tracked back to SFRD differences between these two models (see Fig.~\ref{SFRDOTHERS}).

Similarly, discrepancies in $\bar{I}_{\rm [CII]}$ predicted from models based on the same halo-to-galaxy SFR but different SFR-to-$L_{\rm{[CII]}}$ relations can be understood in light of Eq. \ref{generalSFRL} and Eq.~\ref{ISFR}. Assuming again, that $A\sim1$, we can define two distinct sets of mean SFR-to-$L_{\rm [CII]}$ relations, based on their B value. This way, we have an optimistic set that corresponds to $B \approx 6.9$ of the V15 relation and a pessimistic one that corresponds to $B=6.4$ of the A20 relation. L18 relation, due to its redshift evolution, belongs to the optimistic set at $z=3.7$ and $4.3$ and the pessimistic set at $z=5.8$ and $7.4$. As a result, mean [CII] line intensities predictions from V15 models are optimistic, whereas predictions from the A20 models are pessimistic through the whole redshift range. Predictions of L18 models are close to the ones of V15 at $z=3.7$ and $z=4.3$ and approach the forecasts of A20 models at higher redshifts.

We also tested the influence of $3 \times 10^{9} - 10^{10}\,M_{\odot}$ halos on the forecasted $\bar{I}_{\rm[CII]}$ of the two techniques. For our AM approach, we repeated the procedure with a lower halo mass limit of $3 \times 10^9\,M_{\odot}$. We find that the contribution of low mass halos is only significant for the L18 relation, as it is the only one that predicts bright [CII] luminosities for low SF galaxies ($\text{SFR}<1\,M_{\odot}\,yr^{-1}$; Fig.~\ref{fig:LCII}). However, even in this case, the mean [CII] intensity predicted by the AM method only increases by a factor 1.1 and 1.5 at $z=5.8$ and $z=7$, respectively (Fig.~\ref{fig:PIX}). To test the influence of $3 \times 10^{9} - 10^{10}\,M_{\odot}$ halos for our rTNG approach, we calculated their mean [CII] intensity within the high-resolution TNG100-1 simulation at the central redshift of our FYST spectral window (Fig.~\ref{fig:PIX}). Again, we find that the contribution of these low mass halos is only significant in the case of the L18 relation, with only an increase of the mean [CII] intensity by a factor 1.3 and 2.5 at $z=5.8$ and $z=7$, respectively. A difference is notable only in the two higher-redshift tomographic scans because at higher redshifts, the contribution to the global SFRD of the low-SFR galaxies ($SFR<1\,M_{\odot}\,yr^{-1}$) hosted mainly in low-mass DM halos ($M_{\rm h}<10\,M_{\odot}$) is greater (the trend is visible in Fig.~\ref{fig:LCII}).

In Fig.~\ref{fig:PIX}, we also compare our results to those from \cite{Yue:Ferrara2015} and \cite{Chung:Viero2018}. \cite{Yue:Ferrara2015} use the \cite{Bouwens:Illingworth:2015} UV luminosity function to perform AM and the V15 SFR-to-$L_{\rm{[CII]}}$ relation. Their results are thus naturally in excellent agreement with our AM-V15 predictions. \cite{Chung:Viero2018} use instead the UNIVERSE Machine forward modeling in combination with the L18 relation. At all redshift, their results systematically agree with our most pessimistic predictions. This is true even at low redshift, where L18 corresponds to the most optimistic SFR-to-$L_{\rm{[CII]}}$ relation. \\

Although the [CII] LIM foreground contamination is out of the scope of this paper, we expect that the subtraction of the smooth, IR continuum foreground will largely suppress the [CII] mean intensity, leaving only the intensity fluctuations around the mean \citep{Breysse}. It is for this reason, that we do not consider the possibility of a direct mean intensity measurement.

\subsection{The three-dimensional power spectrum of the [CII] line} \label{PS}

In this section, we use the three-dimensional spherically-averaged power PS to measure the spatial fluctuations in our [CII] tomographic scans and characterize their dependencies with respect to the underlining halo-to-galaxy SFR and the SFR-to-$L_{\rm [CII]}$ relation. We demonstrate in particular how such measurements, which represent the spatial distribution of galaxies weighted by their luminosities, can be used to statistically constrain the halo-to-galaxy SFR and the SFR-to-$L_{\rm [CII]}$ relation without the help of any auxiliary data. 

\subsubsection{Forecasts} \label{FOREC}

To calculate the three-dimensional spherically-averaged PS, we converted our tomographic scans from the angular-frequency space to the three-dimensional comoving space and from there to the Fourier space by computing their Fourier transform. The minimum and maximum scales of the PS accessible to our analysis naturally depends on the "observing" properties of our tomographic scans. For example, the largest physical scale along the line-of-sight (i.e., $r_{\parallel,\rm{max}}$) is defined using the highest and lowest redshifts (i.e., $z_{\rm max}$ and $z_{\rm min}$, respectively) probed by our spectrometer, whereas the smallest scale along the line-of-sight (i.e., $r_{\parallel,\rm{min}}$) is defined by the redshift of two consecutive channels ($z_{\rm{chn},i}$ and $z_{\rm{chn},i+1}$), that is,
\begin{equation}
    r_{\parallel,\rm{max}}=\frac{c}{H_0} \int^{z_{\rm{max}}}_{z_{\rm{min}}} \frac{dz}{\sqrt{\Omega_{\rm{m}}(1+z)^3+ \Omega _{\Lambda}}}\ ,
\end{equation}
\begin{equation}
    r_{\parallel,\rm{min}}=\frac{c}{H_0} \int^{z_{\rm{chn},i+1}}_{z_{\rm{chn},i}} \frac{dz}{\sqrt{\Omega_{\rm{m}}(1+z)^3+ \Omega _{\Lambda}}}\ ,
\end{equation}
where $c$ is the speed of light.
In the plane of the sky, the largest and smallest scales (i.e., $r_{\perp,\rm{max}}$ and $r_{\perp,\rm{min}}$, respectively) probed by our tomographic scans are instead given by, 
\begin{equation}
r_{\perp,\rm{max}}=D_{A}(z_{\rm cen})\ \Delta \theta_S\ ,
\end{equation}
\begin{equation}
r_{\perp,\rm{min}}=D_{\rm{A}}(z_{\rm cen})\ \Delta \theta_{\rm{b}}\ ,
\end{equation}
where $D_{\rm{A}}(z_{\rm cen})$ is the comoving angular distance to the central redshift of our tomographic scans (i.e., $z_{\rm cen}$), $\Delta \theta_S$ is the solid angle covered by our survey in radians, and $\Delta \theta_{\rm{b}}$ is the FWHM of our telescope beam also in radians. Moving to the Fourier space, the largest and smallest scales in the comoving coordinates (from now on, represented by wavenumbers $k$ in units of $\rm{Mpc}^{-1}$; $k_{\parallel}$ and $k_{\perp}$ for the line-of-sight and sky plane scales respectively) become,
\begin{equation}
k_{\rm{min}}=\frac{2 \pi}{r_{\rm{max}}}\ ,
\end{equation}
\begin{equation}
k_{\rm{max}}=\frac{2 \pi}{2 r_{\rm{min}}}\ .
\end{equation}
For the LIM FYST survey, this translates in $k_{\parallel} \in [10^{-3}\,\rm{Mpc}^{-1},10^{-1}\,\rm{Mpc}^{-1}]$ and $k_{\perp} \in [10^{-2}\,\rm{Mpc}^{-1},2\,\rm{Mpc}^{-1}]$ for all our tomographic scans. The scales $k \in [10^{-2}\,\rm{Mpc}^{-1},10^{-1}\,\rm{Mpc}^{-1}]$ are thus available in three dimensions, whereas $k \in [10^{-1}\,\rm{Mpc}^{-1},2\,\rm{Mpc}^{-1}]$ are only available in two dimensions. The importance of this three to two dimensions transition for our sensitivity estimates is discussed in Sect.~\ref{sensitivity}. 

Having defined the limits of the Fourier space accessible by our tomographic scans, their spherically-averaged PS is given by
\begin{equation}
    P(k)=\frac{\langle \tilde{I}_{\rm{[CII]}}^2(k) \rangle}{2 \pi ^2 V_{\rm{box}}}\ ,
\end{equation}
where $\tilde{I}_{\rm{[CII]}}(k)$ is the Fourier transform of these [CII] tomographic scans and $V_{\rm{box}} = r_{\perp,\rm{max}} ^2 \times r_{\parallel,\rm{max}}$ is their volume in comoving units. We performed this calculation using a Fast Fourier Transform algorithm, following \cite{FFT}. 

 \begin{figure*}
    \centering
        \includegraphics[width=1\textwidth]{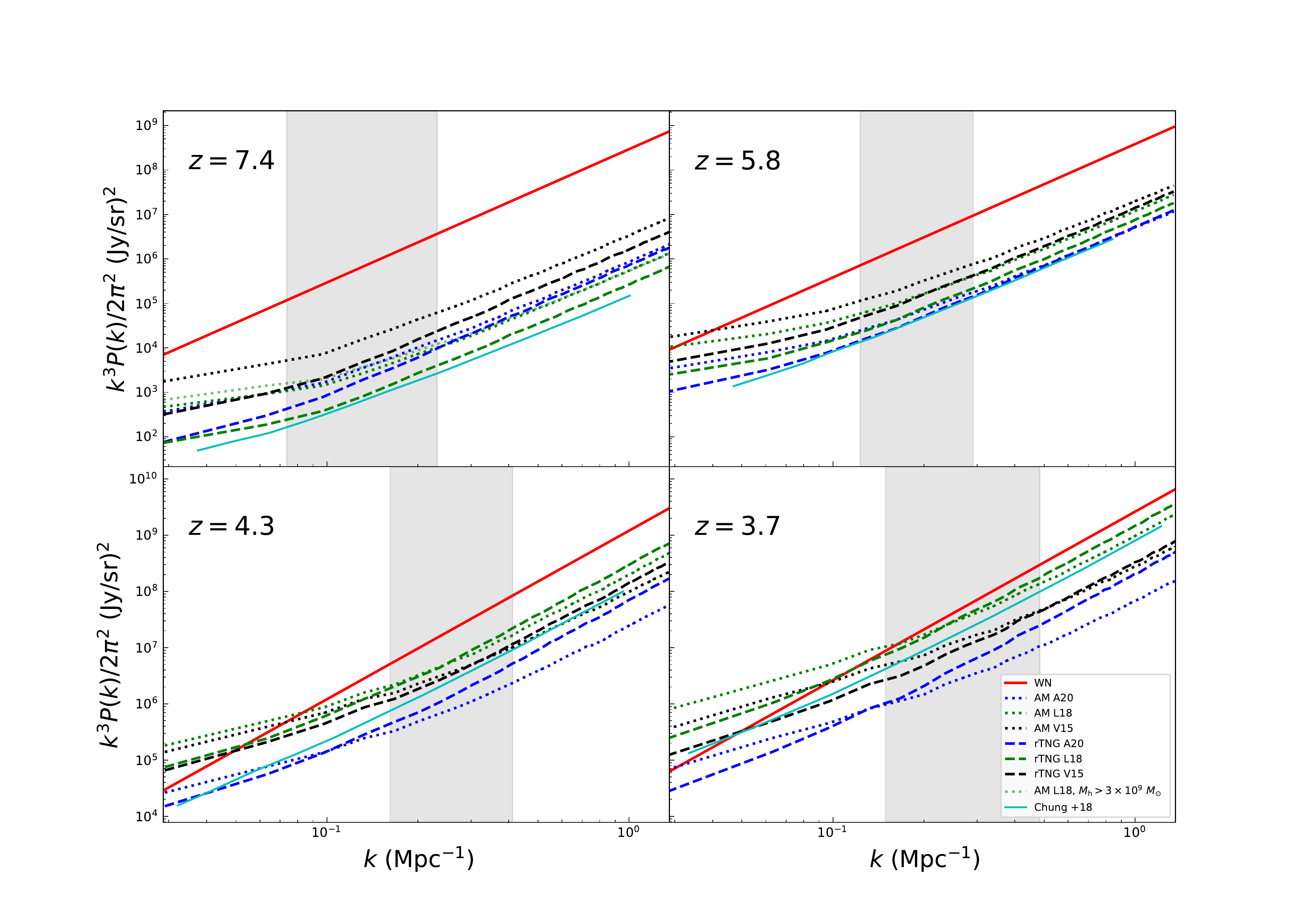}
        \caption{ Spherically averaged PS for a $4^{\circ}\times4^{\circ}$ [CII] mock survey subdivided in four tomographic scans of 40\,GHz bandwidth each centered at $z=3.7$, 4.3, 5.8, and 7.4. The averaging takes place in the Fourier space in $k$-bins of $\Delta k = 0.034\,\rm{Mpc}^{-1}$. Predictions from our rTNG and AM models are shown by dashed and dotted lines, respectively. Lines are color-coded according to the used SFR-to-$L_{\rm{[CII]}}$ relation, blue lines for A20, green lines for L18, and black lines for V15. L18 is also plotted at $z=7.4$ including the contribution of $3 \times 10^{9}-10^{10} \, M_{\odot}$ halos for our AM model (faint green dotted line). The red lines are the power spectrum of the instrumental single-$k$-mode white noise (labeled as WN), $P_{\rm WN}$, of the scheduled FYST LIM survey. Gray areas cover the scales at which $P_{\rm{[CII]}}^{\mathrm{clust}}\sim P_{\rm{[CII]}}^{\mathrm{shot}}$ for our various models. This illustrates the transition from clustering-dominated to shot noise-dominated scales. Predictions from \citet[][turquoise line]{Chung:Viero2018} are also shown for comparison.}
        \label{ALLPS}
\end{figure*}

Fig.~\ref{ALLPS} presents the PS of all the versions of mock [CII] tomographic scans as $k^3P_{\rm{[CII]}}/ (2 \pi ^2)$, in units of $\rm{(Jy/sr)}^2$. All models exhibit the same $\propto k^3$ linear trend at $k>0.3\,\rm{Mpc}^{-1}$ but gradually deviate upward at $k\sim0.1-0.3\,\rm{Mpc}^{-1}$, with this deviation taking place at progressively smaller $k$ as we move to higher redshifts. Despite these similarities, at a given redshift, there is up to two orders of magnitude offset between predictions from different models. Forecasts based on the same SFR-to-$L_{\rm{[CII]}}$ relation but different halo-to-galaxy SFR relation differ by a factor $1-3$ at $k>0.2\,\rm{Mpc}^{-1}$ and $2-6$ at $k<0.2\,\rm{Mpc}^{-1}$, without any systematic evolution of these offsets with redshift. For example, the rTNG models yield more optimistic values for $k>0.2\,\rm{Mpc}^{-1}$ at $z=3.7$ and 4.3, while for the remaining redshifts and scales, the AM models produce systematically higher values. An important difference between the rTNG and AM models is that the deviation from the $\propto k^3$ linear trend is always more significant in the case of the AM models.

Models that are based on the same halo-to-galaxy SFR but different SFR-to-$L_{\rm{[CII]}}$ relations also exhibit differences: across all redshifts, V15 combined with rTNG (AM) yields a factor of two (four) higher PS than A20 at $k>0.2\,\rm{Mpc}^{-1}$ and four (five) at $k<0.2\,\rm{Mpc}^{-1}$; L18, the only relation with a significant redshift evolution goes from being four and two times higher than V15 at $k>0.2\,\rm{Mpc}^{-1}$ and $k<0.2\,\rm{Mpc}^{-1}$) at $z=3.7$, respectively, to four and six times lower than V15 at $k>0.2\,\rm{Mpc}^{-1}$ and $k<0.2\,\rm{Mpc}^{-1}$ at $z=7.4$ (for both rTNG and AM).

As for the mean [CII] intensity, we repeated our AM calculation using a lower DM halo mass limit of $3 \times 10^9\, M_{\odot}$. We found that the contribution of low mass halos to the [CII] PS is only significant for the L18 relation at $z=7.4$, with an increase of the PS by a factor 1.6 at $k_{\rm min} \approx 10^{-2}\,\rm{Mpc}^{-1}$ (Fig.~\ref{ALLPS}). As described in the next paragraph, this value agrees with the square of the $\times1.3$ amplification of the mean [CII] intensity due to these low mass DM halos and inferred in Sect.~\ref{mean intensity}. Unfortunately, we cannot simply calculate the contribution of $3 \times 10^{9}-10^{10} \, M_{\odot}$ halos to the [CII] PS predicted from our rTNG models because the high-resolution TNG100-1 simulation does not probe the necessary large volumes. Nevertheless, we can infer from our mean [CII] intensity analysis that their contribution should be significant for L18 at $z=5.8$ and $z=7.4$, with an increase of the PS signal at low $k$ by a factor $1.3^2=1.69$ and $2.5^2=6.25$, respectively. These increases would put our rTNG predictions at roughly the same levels as those from our AM approach. 

To better understand the dependencies of the PS on the halo-to-galaxy SFR relation and SFR-to-$L_{\rm{[CII]}}$ relation, we study the analytical form of its components. One of these component is the so-called Poissonian shot noise ($P_{\rm{[CII]}}^{\text {shot }}$) arising from the discrete nature of galaxies. Following \citet{Uzgil2014}, the analytical form of $P_{\rm{[CII]}}^{\text {shot }}$ can be written as 
\begin{equation} \label{eq:secondmoment}
P_{\rm{[CII]}}^{\text {shot}}=\int  \frac{dn}{d \log L_{\mathrm{[CII]}}} \left(\frac{L_{\rm{[CII]}}}{4 \pi D_{L}^{2}}y_{\rm{[CII]}} D_{A}^{2}\right)^{2} d \log L_{\mathrm{[CII]}}\ ,
\end{equation}
which, assuming a simple form for the SFR-to-$L_{\rm [CII]}$ relation (Eq.~\ref{generalSFRL}), yields
\begin{equation} \label{eq:secondsecondmoment}
P_{\rm{[CII]}}^{\text {shot}}=10^{2B} \int \frac{dn}{d \log(\text{SFR})} \left(\frac{\text{SFR}^{A} 10^{\sigma _{\rm{L}} ^2 /2}}{4 \pi D_{L}^{2}}y_{\rm{[CII]}} D_{A}^{2}\right)^{2} d \log ( \text{SFR})\ .
\end{equation}
Shot noise is by definition a scale-independent effect and as a result $ P_{\rm{[CII]}}^{\text {shot }}$ is not a function of $k$. In units of $\rm{(Jy/sr)}^2$, this corresponds to the $\propto k^3$ linear trend observed at $k>0.2\,\rm{Mpc}^{-1} $ for all models and all redshifts in Fig.~\ref{ALLPS}. In addition, the fact that $P_{\rm{[CII]}}^{\text {shot}}$ is proportional to SFR$^2$ rather than SFR (assuming $A \approx 1$) explains why the rTNG models yield higher PS than AM models at $k>0.2\,\rm{Mpc}^{-1}$ at $z=3.7$ and 4.3, despite having significantly lower SFRD (Fig.~\ref{SFRDOTHERS}). Indeed, this SFR$^2$ dependency renders the shot noise very sensitive to galaxies with high SFRs (i.e., galaxies with $\text{SFR}>1\,M_{\odot}\,\rm{yr}^{-1}$) and which are more abundant in the rTNG models than in our AM models.

The second component of the PS is the so-called clustering signal component arising from the fact that galaxies follow the dark matter density field. Following again \citet{Uzgil2014}, this component is analytically given by,
 \begin{eqnarray} \label{eq:Pcl}
\begin{aligned} 
&P_{\rm{[CII]}}^{\mathrm{clust}}(k) =\bar{I}_{\rm{[CII]}}^{2} \bar{b}_{\rm{[CII]}}^{2} P_{\delta \delta}(k)\\ &= \left( \int    \frac{dn}{d \log L_{\mathrm{[CII]}}} \frac{L_{\rm{[CII]}}}{4 \pi D_{L}^{2}} y_{\rm{[CII]}} D_{A}^{2} d \log L_{\mathrm{[CII]}} \right) ^{2} \bar{b}_{\rm{[CII]}}^{2} P_{\delta \delta}(k),
\end{aligned}
\end{eqnarray}
which, using Eq.~\ref{generalSFRL}, yields 
\begin{eqnarray} \label{eq:PclSFR}
\begin{aligned} 
&P_{\rm{[CII]}}^{\mathrm{clust}}(k) =\bar{b}_{\rm{[CII]}}^{2} P_{\delta \delta}(k) \\ & \times 10^{2B} \left( \int   \frac{dn}{d \log(\text{SFR})} \frac{10^{\sigma _{\rm{L}}^{2}/2} \text{SFR}^{A}}{4 \pi D_{L}^{2}} y_{\rm{[CII]}} D_{A}^{2} d \log ( \text{SFR}) \right) ^{2}
\end{aligned} 
\end{eqnarray}
where $P_{\delta \delta}(k)$ is the PS of the nonlinear matter and $\bar{b}_{\rm{[CII]}}$ is the average galaxy bias weighted by [CII] luminosity of galaxies. $P_{\delta \delta}(k)$ peaks at a scale that is well constrained by observational cosmology, that is, $k_{\rm eq}=0.01034\pm0.00006\,\rm{Mpc}^{-1}$, and that is set by the Hubble scale at the matter-radiation equality which occurs at $z_{\rm eq}=3387\pm21$ \citep{Planck18}. As a result, $P_{\rm{[CII]}}^{\mathrm{clust}}$, contrary to $P^{\rm shot} _{\rm [CII]}$, is a function of $k$ and peaks at $k\sim10^{-2}\,\rm{Mpc}^{-1}$.

Given that $A\sim1$, $\bar{b}_{\rm{[CII]}}\sim1$ and $P_{\delta \delta}(k)\sim1$ at $k\sim10^{-2}\,\rm{Mpc}^{-1}$, one understands by comparing Eq.~\ref{eq:PclSFR} and Eq.~\ref{eq:secondsecondmoment} that the halo-to-galaxy SFR relation controls how much the $k^3P(k)/(2 \pi ^2)$ signal deviates at $k<0.2\, \rm{Mpc}^{-1}$ from the $\propto k^3$ linear trend. Consequently, the fact that the rTNG models have at all redshifts a large fraction of their SFRD produced by $\text{SFR}>1\,M_{\odot}\,yr^{-1}$ galaxies naturally implies lower $P_{\rm{[CII]}} ^{\rm clust}$-to-$P_{\rm{[CII]}} ^{\rm shot}$ ratios than in the case of AM models.

In short, the different influence of the two steps on the large scales comes from the fact that while the halo-to-galaxy SFR models deviate more significantly at low SFRs (Fig.~\ref{fig:SFRD}), the SFR-to-$ L_{\rm[CII]}$ relations are close to parallel in logarithmic scale (Fig.~\ref{fig:LCII}). Consequently, these differences at low SFRs affect more the amplitude of the clustering signal component ($ P_{\rm{[CII]}} ^{\rm clust} \sim \text{SFR}$) than the shot noise component ($ P_{\rm{[CII]}}^{\text {shot}} \sim \text{SFR}^2$). The $ P_{\rm{[CII]}} ^{\rm clust}$-to-$P_{\rm{[CII]}} ^{\rm shot}$ ratio is in that respect an important observational tool to constrain the halo-to-galaxy SFR relation without being too sensitive to the exact form of the SFR-to-$ L_{\rm [CII]}$ relation.

\subsubsection{Comparison to previous work} \label{COMP}
\begin{figure}
    \centering
        \includegraphics[width=0.53 \textwidth]{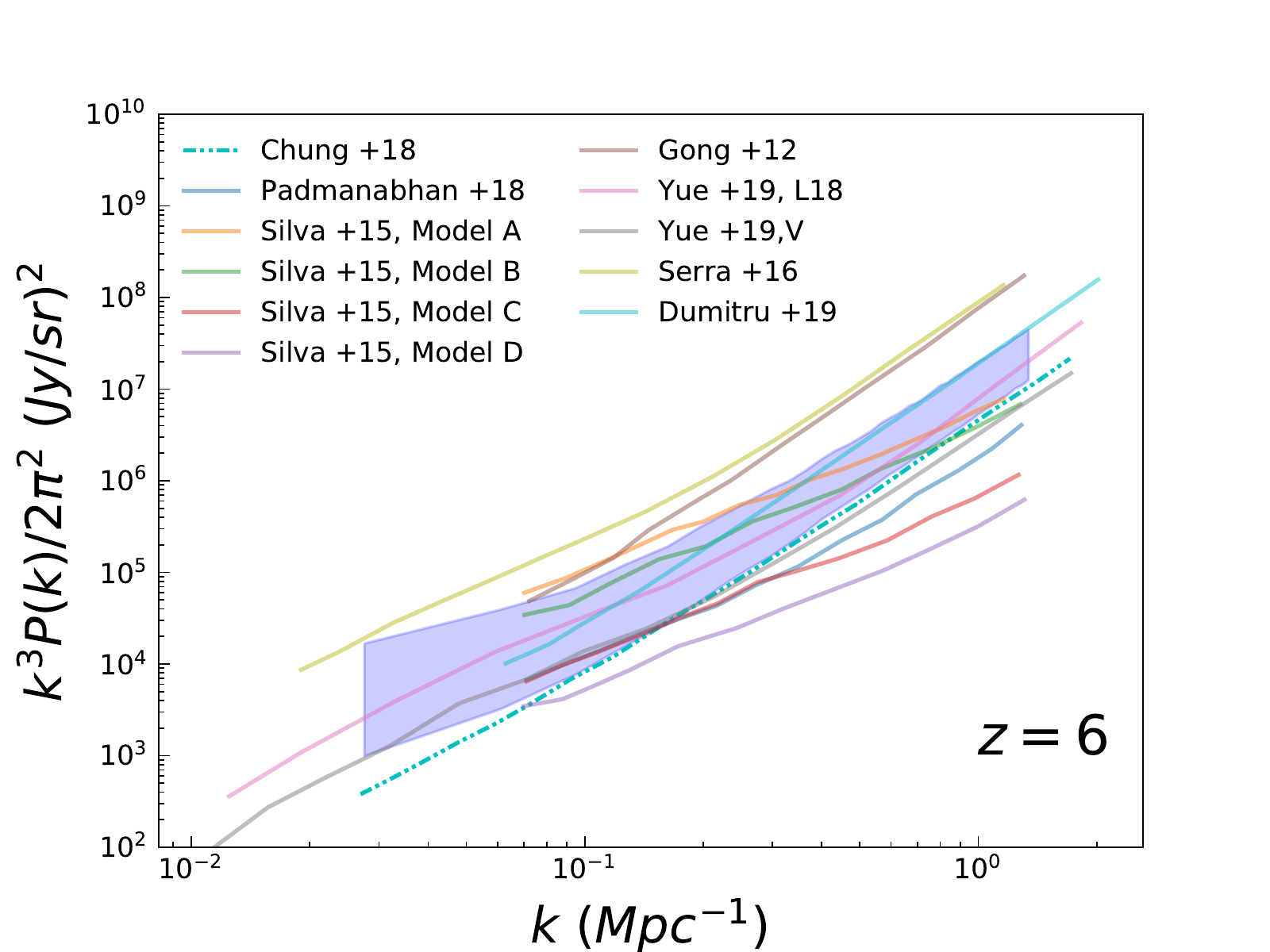}
        \caption{A compilation of $z=6$ [CII] power spectrum predictions from the literature, compared to the range forecasted by our models (shaded area).}
        \label{SHADE}
\end{figure}
Despite the large diversity of models tested here, we could not reproduce the more than two orders of magnitude difference between the most optimistic and pessimistic $P_{\rm{[CII]}}$ forecasts found in the literature (e.g., see Fig.~\ref{SHADE} for $z=5.8$). To investigate the origin of these discrepancies, we examine each case closely, focusing on how their different assumptions influence their forecast.

\cite{Gong:Cooray:2012} do not model the SFR of high-redshift galaxies to predict their $L_{\rm{[CII]}}$, but instead base their $P_{\rm{[CII]}}$ predictions on the average number density of [CII] ions and temperature of the dense high-redshift ISM. This approach is independent of any high-redshift SFRD assumption, but it is highly dependent on the very uncertain fraction of the ISM gas residing in dense clumps. This very different approach yields one order of magnitude higher $P_{\rm{[CII]}}$ values than our most optimistic model. 

\cite{Silva2015} combine a $M_{\rm{h}}$-to-SFR relation from semi-analytic models with four different empirically calibrated SFR-to-$L_{\rm{[CII]}}$ relations with no scatter (label model A, B, C, and D in Fig.~\ref{SHADE}). Changes in the zero points of these four SFR-to-$L_{\rm{[CII]}}$ relations drive most of the offset observed between their models. Two of their models are well in the range of our predictions, but two of them predict much lower PS than ours. A significant difference between \cite{Silva2015} and all models presented in Fig.~\ref{SHADE} is their very shallow slope at $k>0.1\,\rm{Mpc}^{-1}$. This is due to a combination of a strong clustering signal and a weak shot noise signal, explained respectively, by a slight overestimation of the SFRD at $z\sim6$ (see Fig.~\ref{SFRDOTHERS}) and by a flat $M_{ \rm h}$-SFR relation at $M_{ \rm h}>10^{11.5}\,M_{\odot}$.

\cite{Serra2017} base their predictions on an analytic halo model combined with measurements of the cosmic infrared background (CIB). They convert $L_{\rm{IR}}$ into $L_{\rm{[CII]}}$ using empirical relation from $z<4$ galaxies. They initially produced two sets of results: one from the $\it{Planck}$ and one from the $\it{Herschel}$ CIB measurements, their final result being the average of the two sets. The fact that the high-redshift SFRD inferred by \cite{P14} is one order of magnitude higher than that of \citet[][see Fig.~\ref{SFRDOTHERS}]{Madau:Dickinson2014}, results in a strong clustering signal prediction. Combined with a strong shot noise signal due to the high SFR of galaxies hosted in DM halos of $M_{\rm{h}} > 10^{11}\, M_{\odot}$, their result is the most optimistic in our compilation, 0.5\,dex higher than our most optimistic model at $k>0.1\,\rm{Mpc}^{-1}$. Using only the \textit{Herschel} CIB measurements, which are compatible with the \cite{Madau:Dickinson2014} SFRD, would result in one order of magnitude lower forecast well within the range of our predictions.  

\cite{Chung:Viero2018} use the UNIVERSE Machine \citep{Behroozi} to create a $2^{\circ} \times 2^{\circ}$ cone populated with SF galaxies which they translate into mock tomographic scans by adopting the SFR-to-$L_{\rm{[CII]}}$ relation of L18 with a scatter of 0.5\,dex. The UNIVERSE Machine prediction for the SFRD at $z>4$ is half of the \citet[][see Fig.~\ref{SFRDOTHERS}]{Madau:Dickinson2014}, whereas the SFR hosted in DM halos of $M_{h}>10^{11}\, M_{\odot}$ is close to our rTNG models \citep{Behroozi}. The result is a strong shot noise component combined with a weak clustering signal ($k_{\rm tr} \approx 0.1-0.2 \,\rm{Mpc}^{-1}$, close to the $k_{\rm tr}$ of our rTNG L18 model).  

\cite{Dumitru2019} assume that SFR $\propto M_{\rm h}$ and calibrate their model with the cosmic SFRD of \cite{Madau:Dickinson2014}. They use the SFR-to-\(L_{\rm{[CII]}}\) relation of L18 without considering any scatter. The fact that they adopt a linear SFR-$M_{\rm h}$ relation results in a $P_{\rm{[CII]}}$ dominated by its shot noise component, which is well within our prediction range.
\cite{Yue:Ferrara2019} use an analytical halo model calibrated with the SFRD of \cite{Bouwens:Illingworth:2015} combined with several SFR-to-\(L_{\rm[CII]}\) relations. Here we present their forecast based on V15 with a 0.4\,dex scatter and L18 with a 0.6\,dex scatter. Their $P_{\rm{[CII]}}$ predictions are in good agreement with ours. They adopt the $M_{\rm h}$-SFR relation derived from \cite{Yue:Ferrara2015} using AM, resulting in a PS shape similar to our AM models. 

\cite{Padmanabhan2019} combine empirical constraints on the local [CII] line luminosity function with the redshift evolution of the SFRD of \cite{Madau:Dickinson2014}, as well as with the [CII] intensity mapping measurement of \cite{Pullen2018}. Their SFR-to-$L_{\rm [CII]}$ relation is quite different from the rest of the models presented in Fig.~\ref{SHADE}, forecasting weak [CII] emission from the low SFR galaxies ($\text{SFR} < 1\,M_{\odot}\,yr^{-1}$) and making their prediction the most conservative for $z=6$, with the exception of two models of \cite{Silva2015}.

The above comparisons reinforce the conclusion of the previous section. The magnitude of the shot noise component is mostly sensitive to the choice of the SFR-to-$L_{\rm [CII]}$ relation and at a lower level to the shape of the $M_{\rm h}$-SFR relation for $M_{\rm h}>10^{11}\, M_{\odot}$. On the contrary, the magnitude of the clustering component is mostly sensitive to the cosmic SFRD and at a lower level to the choice of the SFR-to-$L_{\rm [CII]}$ relation. The differences between all these forecasts can thus be always traced back to different assumptions on the cosmic SFRD \citep{Serra2017}, the massive end of the $M_{\rm h}$-SFR relation \citep{Silva2015,Dumitru2019}, or the SFR-to-$L_{\rm [CII]}$ relation \citep{Padmanabhan2019}. In the light of the latest observations and simulations, some of these assumptions are outdated or unrealistic. Excluding the \cite{Gong:Cooray:2012} and \cite{Serra2017} models because they significantly overestimate the observed SFRD of \citet{Madau:Dickinson2014} and the \cite{Silva2015} models because they significantly underestimate the SFR hosted in $M_{\rm h}>10^{11.5}\, M_{\odot}$ halos \citep[compared to the latest work on halo-galaxy relation, e.g., ][]{Behroozi}, we end up with PS predictions from the literature consistent with the range observed in our models. This one order of magnitude differences emphasizes the need for more detailed modeling of the star formation and ISM condition of high-redshift galaxies, which should come hand-in-hand with the upcoming LIM observations. 

\subsubsection{Sensitivity estimation} \label{sensitivity}
\begin{figure*}
    \centering
        \includegraphics[width=1\textwidth]{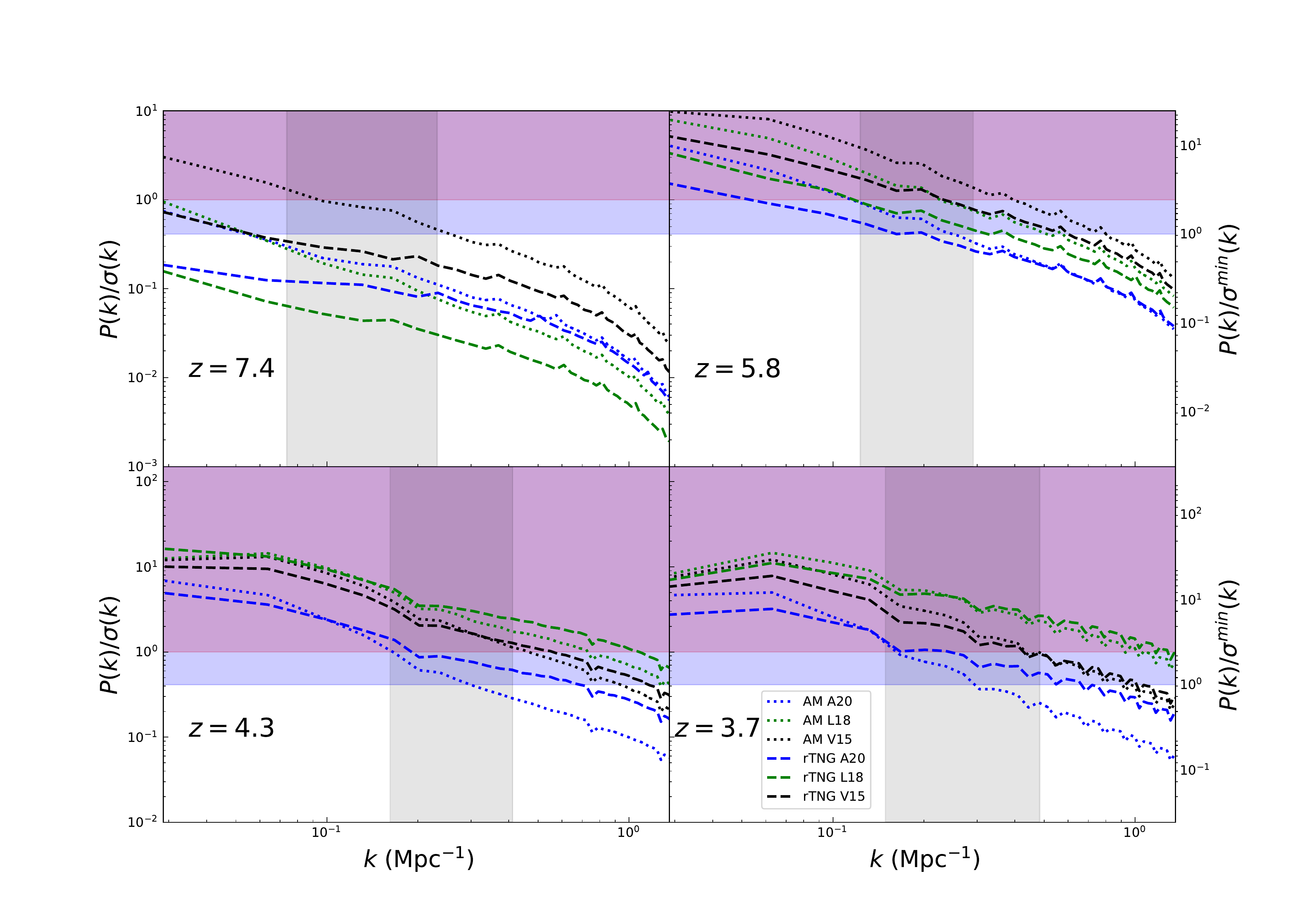}
        \caption{S/N for a $4^{\circ}\times4^{\circ}$ FYST [CII] LIM mock survey of $t_{\rm surv}=4000\,$ hours. It consists of four tomographic scans of 40\,GHz bandwidth each centered at $z=3.7$, 4.3, 5.8, and 7.4. The S/N combines three effects: the instrumental white noise, the sample variance within each $k$-bin, and the attenuation due to smoothing by the instrumental beam. Lines are color-coded according to the used SFR-to-$L_{\rm{[CII]}}$ relation, blue lines for A20, green lines for L18, and black lines for V15. The magenta painted area represents the $S/N>1$ values for the minimum $\Delta k$ considered ($\Delta k = 0.034\,\rm{Mpc}^{-1}$, main $y$-axis), whereas the purple painted area denotes the $S/N>1$ values for the maximum $\Delta k$ considered ($\Delta k=0.34\,\rm{Mpc}^{-1}$, secondary $y$-axis). Gray areas cover all $k$ values for which $P_{\rm{[CII]}}^{\mathrm{clust}}=P_{\rm{[CII]}}^{\mathrm{shot}}$ as indicators for the transition scale between the clustering dominated to the shot noise dominated scales. }
        \label{fig:ALLUN}
\end{figure*}

The signal-to-noise (S/N) achieved when measuring the PS from LIM observations is a combination of three effects: (i) the instrumental white noise, (ii) the sample variance within each $k$-bin, and (iii) the attenuation of the PS signal due to smoothing by the instrumental beam. Based on our $P_{\rm [CII]}$ forecasts, we calculate the S/N for the case of Prime-Cam \citep{vavagiakis/etal:2018}, that is, we assume a telescope diameter of 6\,m, a number of detectors of $N_{\rm beams}=1004$, a total bandwidth per spectrometer of $\Delta \nu =40\,$GHz, a survey covering a $4^{\circ} \times 4^{\circ}$ sky region consisting of $t_{\rm surv}=4000\,$ hours. Here, we do not consider the atmospheric and astronomical foregrounds--methods for the mitigation of which will be investigated in Karoumpis et al. in prep.

Using on-sky noise equivalent intensities, $\sigma_{\rm vox}$, of 0.7, 0.86, 1.7, and 2.8\,MJy\,sr$^{-1}\,$s$^{1/2}$ at 225, 280, 350, and 410\,GHz (CCAT-prime collaboration et al. in prep.) and assuming a homogeneously covered survey, the instrumental white noise can be expressed as,
\begin{equation} \label{pixel spectrum}
    P_{\rm WN} = \frac{\sigma ^{2}_{\rm{vox}}}{t_{\rm{vox}}}V_{\rm{vox}}\ ,
\end{equation}
where $V_{\rm{vox}}$ is the comoving volume covered by a voxel and $t_{\rm vox}$ is the on-sky integration time of this voxel, which is related to the total observing time of our survey (i.e., $t_{\rm surv}$) by,
\begin{equation} \label{surveytopixtime}
    t_{\rm{surv}} = \frac{\text{Number of voxels of the tomography}}{\text{Number of pixels of the detector}} \times t_{\rm{vox}}\ ,
\end{equation}
as the four spectral windows of Prime-Cam are observed simultaneously but only one channel at a time (with the spectral coverage being achieved by adjusting one step at a time the Fabry-Perrot spacing). Each tomography has a different number of voxels, which depends on the angular and spectral resolution of the individual spectral window (see Sect.~\ref{mean intensity}). 

The instrumental white noise is then combined with the predicted PS in order to estimate the statistical uncertainty induced by the finite number of Fourier modes averaged in every $k$-bin,
\begin{equation} \label{eq:uncertainty}
    \sigma _{P(k)} = \frac{P_{\rm [CII]}(k)+P_{\rm{WN}}}{\sqrt{N_{\rm{m}}(k)}}\ ,
\end{equation}
where $N_{\rm{m}}(k)$ is the number of measured modes within a $k$-bin centered at $k$. As discussed in \cite{Chung:Viero2018}, $N_{\rm{m}}(k)$ is given by,  
\begin{equation} \label{Noofmodes}
N_{m}(k)=\frac{\min (k, k_{\parallel , \rm{max}}) \ k \ \Delta k \ V_{\mathrm{surv}}}{4 \pi^{2}}\ ,
\end{equation}
where the term $\min (k, k_{\parallel , \rm{max}})$ accounts for the fact that $k_{\parallel,\rm{max}}$ is an order of magnitude smaller than $k_{\perp,\rm{max}}$, which implies that $k$-bins greater than $k_{\parallel,\rm{max}}$ have their three-dimensional sphere truncated at $k_{\parallel}> k_{\parallel,\rm{max}}$. We note that neglecting the effect of this three to two dimensions transition would result in an overestimation of the S/N at large $k$.

Finally, following \cite{Li:Wechsler2016}, we account in the calculation of the S/N for the attenuation of the PS signal caused by the smoothing of the intensity map by the instrumental beam. This is done using a modification of the attenuation factor $W =P(k)/P_{\rm SM}(k)$ \citep{Li:Wechsler2016}, $P_{\rm SM}(k)$ being the PS of the smoothed map. For the case of asymmetrical voxels, it comes that, 
\begin{equation} \label{beamsmooth}
W ( k ) = e ^ { - k ^ { 2 } \sigma _ { \perp } ^ { 2 } } \int _ { 0 } ^ { 1 } e ^ {-\rm{min}( \mu k, k_{\parallel , max} ) ^{ 2 }\ \left( \sigma _ { \| } ^ { 2 } - \sigma _ { \perp } ^ { 2 } \right) }\ d \mu \ ,
\end{equation}
with,
\begin{equation}
\sigma_{\|}=\frac{c}{H(z)}\ \frac{\Delta\nu_{\rm{b}}\,(1+z)}{2.355 \ \nu_{\mathrm{obs}}}
\end{equation}
and,
\begin{equation}
\sigma_{\perp}=\frac{D_{\rm{A}}(z)\,\Delta \theta _{\rm{b}}}{2.355}
\end{equation}
where $\Delta\nu_{\rm{b}}$ and $\Delta \theta_{\rm{b}}$ are the FWHM of the angular and spectral beams, respectively. 

Bringing together the three effects driving the noise, the instrumental white noise, sample variance, and resolution limits, the S/N can be written as,
\begin{equation} \label{signaltonoise}
\mathrm { S } / \mathrm { N } = W ( k )\ \sqrt { N _ {\rm{m} } ( k ) }\ \frac { P _ { \mathrm { [CII] } } ( k , z ) } { P _ { \mathrm { [CII] } } ( k , z ) + P _ { \mathrm { N } } }\ .
\end{equation}

The S/N achieved by the FYST LIM survey using a $k$-bin of $\Delta k=0.034\,\rm{Mpc}^{-1}$ (i.e., the finest $k$-space resolution of the survey) are shown in Fig.~\ref{fig:ALLUN}. For all models, the achieved S/N decreases significantly with increasing $k$. The offsets in S/N between these models are to first order the same than those observed between their predicted PS (Fig.~\ref{ALLPS}). The only exception is at $k <0.06\,\rm{Mpc}^{-1}$ for $z=3.7$ and 4.3, where most models predict $P_{\rm{[CII]}}>P_{\rm{WN}}$ and thus have S/N which converge toward $\sqrt{N _ {\rm{m}}( k )}$.

At $k<0.2\,\rm{Mpc}^{-1}$, the fall in S/N is caused by the increase of the $P_{\rm{WN}}$-to-$P_{\rm{[CII]}}$ ratio with increasing $k$ (see Eq.~\ref{signaltonoise} and Fig.~\ref{ALLPS}). Then, at $k>0.2\,\rm{Mpc}^{-1}$, the $P_{\rm{WN}}$-to-$P_{\rm{[CII]}}$ ratio remains constant but the attenuation factor $W(k)$--accounting for the influence of the beam--becomes significant, steepening further the fall of the S/N. We note also that the restricted spectral resolution of the Prime-Cam influences the growth of $N(k)$. At $k>k_{\parallel, \rm{max}}$, the line-of-sight modes become indeed unavailable, which causes a discontinuity in the slope of the S/N at $k\sim0.2\,\rm{Mpc}^{-1}$.

Finally, we observe periodic jumps in S/N (or in the case of $z=3.7$ and 4.3 periodic dips) with a different period at different redshift. Those are explained by periodic jumps in the number of Fourier modes averaged in a given $k$-bin. Indeed, because our voxels have a cuboid shape, a shell of pixels whose distances are $k\pm\Delta k/2$ from the center of the tomography, is only an approximation of a spherical shell. For example, when averaging over $k$-bins of $\Delta k =0.034\,\rm{Mpc}^{-1}$, the width of the shells oscillates between three and four pixels at 225 and 280\,GHz and between two and three pixels at 350 and 410\,GHz, respectively. 

The FYST LIM survey will be optimal for constraining at high $k$-resolution the clustering component of the [CII] PS. Indeed, at the scales where this component dominates (i.e., $k<0.2\,\rm{Mpc}^{-1}$), even our most pessimistic models yields clear ${\rm S/N}>1$ detection at all redshifts but $z=7.4$. The detection at such high $k$-resolution of the [CII] PS at scales where the shot noise component dominates (i.e., $k>>0.2\,\rm{Mpc}^{-1}$) will be most challenging for the FYST LIM survey, with only two models detected at $z=3.7$, four at $z=4.3$, one at $z=5.8$, and none at $z=7.4$. This component, which is invariant with $k$, can, however, be fully constrained using only two measurements made at sufficient $k$ leverage. Finally, at $z=7.4$, the FYST LIM survey will only detect the clustering component of the [CII] PS in the case of our most optimistic model.

In order to increase the probability of detecting the [CII] PS at high redshifts and large $k$, it is conceivable to increase the value of $\Delta k$ by up to an order of magnitude, amplifying the S/N by a factor of $\sqrt{10}$ (see right-hand $y$-axis of Fig.~\ref{fig:ALLUN}). This re-binning would result in three data points for each measured PS: one at scales where the clustering component dominates (all models detected at $z=3.7$, 4.3 and 5.8; four models at $z=7$), one at intermediate scales (all models detected at $z=3.7$, 4.3 and 5.8; one model at $z=7.4$), and one at scales where the shot-noise component dominates (five models detected at $z=3.7$ and 4.3; three at $z=5.8$; none at $z=7$). 

\section{Discussion}
\label{discussion}

\begin{figure}[t]
    \centering
        \includegraphics[width=0.53 \textwidth]{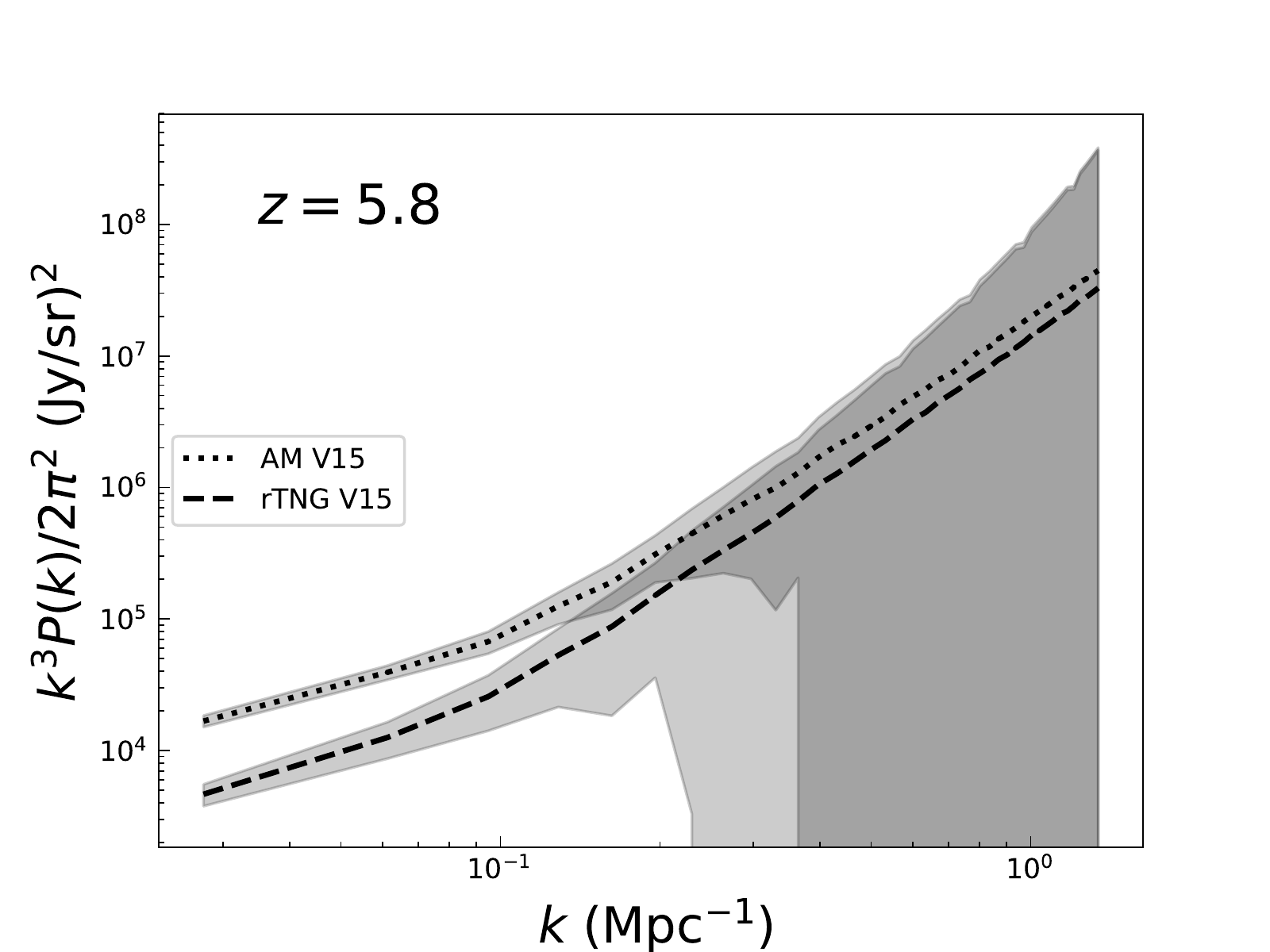}
        \caption{[CII] PS at $z=5.8$ for our AM and rTNG models combined with the SFR-to-$L_{\rm[ CII]}$ relation of V15. Results are the same as in Fig.~\ref{ALLPS} but this time is plotted with the uncertainties coming from the S/N of Fig.~\ref{fig:ALLUN} for $\Delta k= 0.034\,\rm{Mpc}^{-1}$. The plot serves as an example of the potential of FYST to trace the [CII] PS from the EoR on large scales constraining the halo-to-galaxy SFR relation.}
        \label{fig:PSwithError}
\end{figure}

Our study unambiguously demonstrates that the latest physically- or empirically-motivated models describing the halo-to-galaxy SFR relation and the SFR-to-$L_{\rm{[CII]}}$ relation at the (post-)EoR yield [CII] LIM predictions spanning a range of one order of magnitude. Such large uncertainties pose, naturally, a challenge for designing [CII] LIM experiments, as those must be tailored to allow for the detection of even the most pessimistic yet realistic model. On the other hand, these variations also demonstrate the power of future [CII] LIM measurements to restrict the range of possible early galaxy evolution models. In particular, by detecting the [CII] PS at low- and high-spatial scales LIM experiments can put stringent constraints on the yet very uncertain redshift evolution of the SFRD, the SFR-to-$L_{\rm [CII]}$ relation during the (post-)EoR as well as shed light on the halo-to-galaxy SFR relation (Fig.~\ref{fig:PSwithError}).

In this context, FYST will certainly unravel unknown aspects of early galaxy evolution. The observations of FYST will constrain the cosmic SFRD up to $z \sim 7$, a key parameter for modeling the evolution of galaxies over the critical first gigayears of the Universe, the current knowledge of which is hampered by observational limitations. Even though galaxy surveys performed by ALMA and JWST will also attempt to measure the redshift evolution of the SFRD at the (post-)EoR, their optical design is optimal for deep, "pencil-beam" galaxy surveys (e.g., the field of view is $\sim 20$ arcsec for ALMA and $\sim 180$ arcsec for {\em JWST} compared to $\sim 4680$ arcsec of the Prime-Cam spectro-imager of the FYST). These surveys are excellent for studying the properties of individual galaxies but they will provide highly cosmic variance-limited statistical measurements compared to those from the LIM surveys of FYST. In addition, the much wider surveys of FYST will render possible the measurement of the clustering properties of SF galaxies at the (post-)EoR and thus shed light on the formation of the first large scale structure of the Universe.

Naturally, FYST is not the sole experiment that aims at performing [CII] LIM maps of the (post-)EoR. Nevertheless, while the TIME and CONCERTO experiments will rely as FYST on state-of-the-art spectro-imager instruments, they will operate on classic single-dish antennas as opposed to the novel "crossed Dragoned" configuration of FYST. The optical design of their instruments (e.g., field of view $\sim 840$ arcsec for TIME and $\sim 1200$ arcsec for CONCERTO) and their higher spatial- and spectroscopic-resolution motivates them to focus on smaller scales than FYST: the [CII] LIM planned surveys of TIME and CONCERTO are $0.1^{\circ} \times 0.1^{\circ}$ and $1.3 ^{\circ} \times 1.3 ^{\circ}$ wide, respectively, compared to the envisioned $4^{\circ} \times 4^{\circ}$ LIM survey of FYST. The higher resolution of CONCERTO and TIME will allow them to put stringent constraints on the [CII] PS shot-noise signal, which is mostly out of reach to FYST low-resolution observations. On the other hand, the wider survey will allow FYST to be the only [CII] LIM experiment to detect the [CII] PS clustering signal at the EoR and post-EoR \footnote{We note that CONCERTO could also detect the post-EoR [CII] PS clustering signal at $ z \approx 4.5$ \citep[see][]{Chung:Viero2018}.}.
Combining the [CII] PS measurements of the different experiments will therefore result in even more powerful constraints.  

A common challenge for all LIM experiments is foreground contamination. The most critical contaminants of [CII] LIM are the cosmic infrared background (CIB) and the CO rotational lines emitted by foreground galaxies. The CIB is spectrally smooth, confined to the large Fourier scales, and can thus be easily rejected during the [CII] LIM statistical analysis. On the contrary, the CO line foreground contamination is not spectrally smooth and requires the development of complex mitigation methods. Those focus either on retrieving the PS or attempting to reconstruct the individual line maps (see CCAT-prime collab. et al. in prep. for a list of the available methods). In the second part of this paper series, we will test and evaluate several foreground mitigation methods using realistic astrophysical contaminants generated from our rTNG cone catalog.

\section{Conclusions}
In this paper, we forecast the [CII] mean intensity and PS at $z = 3.7$, 4.3, 5.8, and 7.4 and investigate the prospect of measuring it with the Prime-Cam spectro-imager of the FYST. We generate various versions of [CII] tomographic scans basing them on a common DM halo cone created using the DM halo catalogs of Illustris TNG300-1 simulation. We approximate the halo-to-galaxy SFR relation either by adopting the integrated SFR of Illustris TNG300-1 galaxies or the SFR coming from abundance matching the DM halos of Illustris TNG300-1 with dust-corrected UV luminosity function of high-redshift galaxies. We couple these two alternatives with three SFR-to-$L_{\rm [CII]}$ relations. We then estimate the [CII] mean intensity and PS of these various tomographic scans and study its detectability, assuming the technical characteristics of FYST and the planned LIM FYST survey of 4000 hours and $4^{\circ} \times 4^{\circ}$ sky coverage. Our main findings can be summarized as follows.
\begin{enumerate}

\item The forecasted mean [CII] line intensities emitted by \ \ \ \  \ (post-)EoR galaxies significantly decrease as we proceed to lower observing frequencies, corresponding to higher redshifts. Although all of our models follow this general redshift trend, there is up a factor of 10 difference among their predictions.

\item The amplitude of the forecasted [CII] intensity power spectrum ranges up to a factor of 30, depending on the selection of the halo-to-galaxy SFR and the SFR-to-L[CII] relations. The magnitude of the shot noise component of the PS is primarily sensitive to the selection of the SFR-to-$L_{\rm [CII]}$ relation and at a more moderate level to the form of the $M_{\rm h}$-SFR relation. The magnitude of the clustering component of the PS is mainly dependent on the cosmic SFRD of the model and at a lower level to the choice of the SFR-to-$L_{\rm [CII]}$ relation.

\item The mass resolution of the TNG300-1 simulation affects the formation and evolution of galaxies residing in low-mass halos ($ M_{\rm h}<10^{10}\, M_{\odot}$), which were therefore excluded from our rTNG cone catalog. Relying on the alternative method of abundance matching, we estimate that the contribution of $ 3\times10^{9}<M_{\rm h}/M_{\odot}<10^{10}$ halos to the [CII] PS signal is only significant at $ z=6$, with a maximum amplification factor of $ \times 1.69$, and $ z=7$, with a maximum amplification factor of $ \times 6.25$.

\item FYST will be optimal to measure the [CII] PS signal at large, clustering-dominated scales ($k\lesssim 5 \times 10^{-2}\,\rm{Mpc}^{-1}$ ), where even our most pessimistic model is detected at $z=3.7$, 4.3, and 5.8 and four out of six models give tentative detections at $\rm z=7.4$. 

\item Detecting the [CII] PS at small, shot noise-dominated scales ($k \gtrsim0.5\,\rm{Mpc}^{-1}$), where five out of six models are detected at $z=3.7$ and 4.3, three out of six at $z=5.8$, and none at $z=7$, will be more challenging for the relatively low-resolution observations of FYST. 

\item The detection of the [CII] PS at low- and high-spatial scales will constrain the halo-to-galaxy SFR relation, disentangling it from the precise form of the SFR-to-$L_{\rm [CII]}$ relation.
\end{enumerate}

Due to the exceptional location and the novel optical design of the telescope, FYST observations present an unparalleled opportunity for performing a [CII] PS measurement. As a result, its results will unravel key evolutionary properties of galaxies during the (post-)EoR.

\begin{acknowledgements}
We thank Dongwoo Chung, Dominik Riechers, Gordon Stacey,and other members of the CCAT-prime science working group for detailed discussions about EoR-Spec; We are grateful to Toma Badescu, Kevin Harrington, Ana Paola Mikler, Jens Erler, Eric F. Jimenez Andrade, Basilio Solis, Maude Charmetant, Ankur Dev, Enrico Garaldi, Victoria Yankelevich, Joseph Kuruvilla, Cristiano Porciani, Kaustuv moni Basu, Lydia Moser-Fischer, Stefanie Mühle, Reinhold Schaaf, and Eleni Vardoulaki for the helpful discussions. This research was carried out within the Collaborative Research Center 956, sub-projects A1 and C4, funded by the Deutsche Forschungsgemeinschaft (DFG) – project ID 184018867. This research made use of NASA’s Astrophysics Data System Bibliographic Services; Matplotlib \citep{Hunter:2007}; Astropy, a community-developed core Python package for astronomy \citep{Astropy}; PlotDigitizer (http://plotdigitizer.sourceforge.net/).
\end{acknowledgements}


\bibliographystyle{aa} 
\bibliography{adstest}

\newpage

\end{document}